\documentclass[aps,prd,12pt,nofootinbib]{revtex4}

\usepackage{graphics}

\usepackage{epsfig}
\usepackage{amsmath}	%American Mathematical Society math package
\usepackage{amssymb}    %American Mathematical Society symbol package

\newcommand{\be}{\begin{eqnarray}}
 \newcommand{\ee}{\end{eqnarray}}
 \newcommand{\non}{\nonumber\\}

 \newcommand{\eqn}[1]{Eq.~(\ref{#1})}
 \newcommand{\bra}[1]{\left\langle #1 \right|}
 \newcommand{\ket}[1]{\left| #1 \right\rangle}

\begin{document}

%%%%%%%%%%%%%%%%%%%%%%%%%%%%%%%%%%%%%%%%%%%%%
% Title, authors, affiliations and abstract %
%%%%%%%%%%%%%%%%%%%%%%%%%%%%%%%%%%%%%%%%%%%%%

\pagestyle{empty}

\title{Tensor Meson Production in Proton-Proton Collisions from the Color Glass Condensate}

%\author
%{Fran\c{c}ois Fillion-Gourdeau, Jean-S\'{e}bastien Gagnon, Sangyong
%Jeon}
%\affiliation
%{Department of Physics, McGill University, 3600 University Street,
%    Montreal, Canada H3A 2T8}

\author{Fran\c{c}ois Fillion-Gourdeau}
\email{ffillion@hep.physics.mcgill.ca}

\author{Sangyong Jeon}
\email{jeon@physics.mcgill.ca}

\affiliation{Department of Physics, McGill University, 3600 University Street, Montreal, Canada H3A 2T8}

\date{\today}

\begin{abstract}
We compute the inclusive cross-section of $f_{2}$ tensor mesons production in proton-proton collisions at high-energy. We use an effective theory inspired from the tensor meson dominance hypothesis that couples gluons to $f_{2}$ mesons. We compute the differential cross-section in the $k_{\perp}$-factorization and in the Color Glass Condensate formalism in the low density regime. We show that the two formalisms are equivalent for this specific observable. Finally, we study the phenomenology of $f_{2}$ mesons by comparing theoretical predictions of different parameterizations of the unintegrated gluon distribution function. We find that $f_{2}$-meson production is another observable that can be used to put constraints on these distributions.  
\end{abstract}

\maketitle

%\pacs{}

\pagestyle{plain}

%%%%%%%%%%%%%%%%%%%%%%%%%%%%%%%
% Main body of the manuscript %
%%%%%%%%%%%%%%%%%%%%%%%%%%%%%%%

\section{Introduction}

High energy hadronic collisions are complex phenomena that combine many-particles physics and Quantum Chromodynamics (QCD). Making predictions for the production of particles in these reactions involves the understanding of both hadron wave-function and processes of particle creation. Many advances have been made in the last decades in those fields through the application of perturbative QCD (pQCD) to the description of experimental data. The pQCD analysis relies generally on factorization schemes like Collinear Factorization and $k_{\perp}$-factorization.  Different procedures are implemented in these approaches to improve the perturbation expansion by resumming infrared divergences. Many observables are well-described by QCD factorizations. For example, $k_{\perp}$-factorization was used successfully in proton-antiproton collisions for heavy-quarks production  \cite{Collins:1991ty,Catani:1990eg,Gribov:1984tu,Kuraev:1977fs,Jung:2001rp,Lipatov:2001ny,Zotov:2003cb,Lipatov:2005at,Andersson:2002cf} while collinear factorization is one of the main computational tool for deep inelastic scattering and for a number of other applications \cite{Brock:1993sz,Jalilian-Marian:2001bu}. The main difference between the two formalisms is their validity range. $k_{\perp}$-factorization describes semihard processes characterized by typical momentum transfer $\mu$ obeying $\Lambda_{QCD}^{2} \ll \mu^{2} \ll s$ while hard processes with $\Lambda_{QCD}^{2} \ll \mu^{2} \sim s$ is the domain of collinear factorization.

Some time ago, new ideas were developed to take into account effects due to the recombination of gluons in nuclei at very high energy (or small-x) \cite{McLerran:1993ka,McLerran:1993ni,Iancu:2002xk,Iancu:2003xm,Venugopalan:2004dj}. These saturation effects introduce a new scale called the saturation scale $Q_{s}$ at which the probability of gluon recombination becomes important. A naive estimation of $Q_{s}$ shows that it depends on the momentum fraction $x$ and the number of nucleon like $Q_{s}^{2} \sim A^{\delta} x^{-\lambda}$ \cite{Iancu:2002xk,Iancu:2003xm}. At small enough $x$ or large enough $A$, the saturation scale becomes hard ($Q_{s}^{2} \gg \Lambda_{QCD}^{2}$) and weak coupling techniques can be used. The Color Glass Condensate (CGC) includes these saturation effects in a semi-classical formalism \cite{McLerran:1993ka,McLerran:1993ni,Iancu:2002xk,Iancu:2003xm,Venugopalan:2004dj}. More precisely, it resums the large order contributions coming from the interaction of gluon cascades in the regime where $\Lambda_{QCD}^{2} \ll \mu^{2} \leq Q_{s}^{2} \ll s$. In proton-proton (pp) collisions at RHIC energy (with $Q_{s} < \mu$), saturation can be neglected and collinear or $k_{\perp}$-factorization can be used in the calculation of observables. It is interesting to compare the predictions of CGC and factorizations in that regime. For heavy quark production \cite{Gelis:2003vh} and gluon production \cite{Gyulassy:1997vt,Kovchegov:1997ke}, the CGC in the low density limit and $k_{\perp}$-factorization are equivalent in pp collisions. We show in this paper that this also hold true for tensor meson production.

In this study, we focus on the production of $f_{2}(1270)$, the lightest spin-2 meson. It is a bound state of light quarks (u or d) and antiquarks, it has a mass of 1.27 GeV, a decay width of 185 MeV and quantum numbers of $I^{G}(J^{PC})=0^{+}(2^{++})$ \cite{Yao:2006px}. This particle was studied in the past using different methods such as chiral perturbation theory \cite{sjrey1,sjrey2,Toublan:1996bk} and effective field theory \cite{Giacosa:2005bw}. An other approach is to use the tensor meson dominance (TMD) hypothesis where one assumes that the energy-momentum tensor form factor are dominated by the exchange of an $f_{2}$ meson. The idea of TMD was first applied to phenomenology by Renner \cite{renner70,renner71} to describe the tensor-meson exchange channel in pion-nucleon scattering. This idea was then used in many other applications \cite{Gampp:1978ym,Gampp:1978xc,Gampp:1979hp,Bongardt:1980qf,Genz:1982yn,Ishikawa:1988xi,Terazawa:1990es, suzuki,Yan:1996xq}.  Borrowing ideas from the TMD hypothesis, namely the coupling of spin-2 mesons to the energy-momentum tensor of strongly interacting matter (including quarks, gluons and hadrons), we show how the inclusive cross-section of $f_{2}$ mesons can be computed in proton-proton collision at high-energy (RHIC and LHC). This type of coupling is used in \cite{suzuki} and in \cite{Katz:2005ir} for the computation of the $f_{2}$ decay rate into photons and pions. In the study described in \cite{Katz:2005ir}, the authors compute successfully (within experimental error) the decay rate of $f_{2}$ into two photons using ADS/QCD. To obtain this accurate result, they must include both the gluon and quark contributions to the energy-momentum tensor. In this paper, we consider only the gluonic part of the energy-momentum tensor since the hadrons in high-energy collisions interact mostly through gluons. More precisely, we study the process where the interaction of two gluons coming from different nuclei interact and produce one on-shell $f_{2}$ meson because this is the dominant process at high-energy. 

We consider only the case of proton-proton (pp) collisions and discuss briefly the case of proton-nuclei (pA) collisions. For nuclei-nuclei (AA) collisions, the total number of $f_{2}$-mesons produced should be important but most of them cannot be detected because they decay inside the medium created by the collision. This can be seen as follow.  Our analysis shows that tensor mesons are created during the first instants of the collision ($t<1 \; \mbox{fm/c}$), when the system is still in non-equilibrium and well described by strong classical field. RHIC data indicates  that from $1 \; \mbox{fm/c}$ up to $10 \; \mbox{fm/c}$, a thermalized medium is created. The mean lifetime of $f_{2}$ mesons is about $1.1 \; \mbox{fm/c}$, which is smaller than the lifetime of the medium.  Moreover, the probability that these particles travel distances larger than the radius of the nuclei (or larger than the medium) is very small unless their momenta obey $|\mathbf{p}| \gg M  $. This is very unlikely because the typical transverse momenta will be $M \approx 1.27 \; \mbox{GeV}$ and the $f_{2}$ are created by small-x gluons which have very small longitudinal momenta to begin with. Thus, the $f_{2}$ will mostly decay inside the medium and the decay products will rescatter, losing information about their origin. In that sense, most tensor mesons produced should be unobservables in AA-collisions unless they are produced on the surface. This can be seen experimentally in the invariant mass distribution where the $f_{2}$ signal is essentially non-existent \cite{Adams:2003cc}. This is not the case for pp and pA because no medium is created and the particles obtained from $f_{2}$ decay can escape without rescattering. For these reasons, our present analysis can only be applied to pp and pA collisions.

This paper is organized as follow. Section 2 is dedicated to the statement of the spin-2 tensor mesons effective theory. We use an effective free Lagrangian describing spin-2 particles dynamics that is similar to Kaluza-Klein modes in extra-dimensions studies. The $f_{2}$ interacts with strongly interacting matter through the  energy-momentum tensor.  In section 3, the computation of the cross-section at leading order in perturbation theory using $k_{\perp}$-factorization is performed. The process considered is the tree-level interaction of two gluons giving one $f_{2}$. In section 4, the same calculation is done using the CGC. We start by deriving a reduction formula relating the cross-section production to a correlator of energy-momentum tensors that can be evaluated using the CGC formalism. We use the solution of the gauge field in covariant gauge at leading order in the color charge densities to evaluate the cross-section.  The two formalisms are then compared and shown to be equivalent in the limit of low densities (dilute systems). In both cases, the production cross-section is related to the unintegrated gluon distribution functions of the two protons. Section 5 is devoted to the phenomenology of $f_{2}$ meson production at RHIC. In that section, the predictions of standard parameterizations of the unintegrated distribution function are compared against each other. 

Throughout the paper, we use both light-cone coordinates defined by
\begin{eqnarray}
p^{+} &=& \frac{p^{0} + p^{3}}{\sqrt{2}} \; ; \; 
p^{-} = \frac{p^{0} - p^{3}}{\sqrt{2}}
\end{eqnarray}
and Minkowski coordinates. It should be clear by the context which one is used. We also use the metric convention $\eta_{\mu \nu} = (1,-1,-1,-1)$.

\section{Effective Theory} 

In our approach, the tensor mesons couples to the rest of the strongly interacting matter through the energy-momentum tensor. The energy-momentum tensor is a conserved current that has the same quantum numbers as tensor mesons, so a coupling of this type is natural. The effective interaction Lagrangian describing the interaction of gluons, quarks and hadrons with tensor mesons is simply given by \cite{suzuki,Katz:2005ir}
\begin{eqnarray}
\label{eq:int_lag}
\mathcal{L}_{\rm{int}}(x) = \frac{1}{\kappa}f_{\mu\nu}(x)T_{\rm{gluons}}^{\mu\nu}(x) + \frac{1}{\kappa_{\rm{q}}}f_{\mu\nu}(x)T_{\rm{quarks}}^{\mu\nu}(x) + \frac{1}{\kappa_{\rm{h}}}f_{\mu\nu}(x)T_{\rm{hadrons}}^{\mu\nu}(x)
\end{eqnarray}
where $f_{\mu\nu}$ is the symmetric spin-2 tensor field, $\kappa$ and $\kappa_{q,h}$ are the coupling constants and $T^{\mu\nu}_{\rm{gluons},\rm{quarks},\rm{hadrons}}$ are the energy-momentum tensors of gluons, quarks and hadrons respectively. This type of interaction preserves all the symmetries of the matter Lagrangian by construction. For simplicity, we consider only one species of tensor meson, namely $f_2(1270)$. Generalization to include other species like the $f_{2}'(1525)$ is straightforward. Note also that in this theory, the coupling constants $\kappa$ are free dimensionful (dimension of energy) parameters that need to be fixed by experiments. It is possible to fix their value by assuming that they are the same ($\kappa=\kappa_{\rm{q}}=\kappa_{\rm{h}}$). Within this assumption, the hadronic and partonic sectors of the theory couple to tensor meson in the same way. Then, $\kappa$ can be fixed using the $f_{2}$ pions decay and one finds that $\kappa \approx 0.1 \;\mbox{GeV}$ \cite{suzuki,Katz:2005ir,Bongardt:1980qf}. The theory used in this paper differs from this approach, we use different coupling constants for each sectors ($\kappa \neq \kappa_{\rm{q}} \neq \kappa_{\rm{h}}$). The value of the coupling is then evaluated by comparing to experiments (see the last section for a comparison with STAR data).

For a physically consistent formulation of $f_2$ tensor Lagrangian, we borrow from recent developments of Kaluza-Klein theory and write the following free Lagrangian describing spin-2 particles dynamics \cite{Giudice:1999ck,Han:1999sg}.
 \be
 {\cal L_{\rm{free}}} (x)
 & = &
 -\frac{1}{2} f^{\mu\nu}(x)(\partial^2 + M^2) f_{\mu\nu}(x)
 +\frac{1}{2} f_\rho^\rho (x) (\partial^2 + M^2) f_\sigma^\sigma (x)
 \non
 & & {} 
 -f^{\mu\nu} (x) \partial_\mu\partial_\nu f_\rho^\rho (x)
 +f^{\mu\nu} (x) \partial_\mu\partial_\sigma f^\sigma_\nu (x)
 \label{eq:L_KK}
 \ee
The first term of the Lagrangian is the usual kinetic term while the other terms are necessary to have the right number of degrees of freedom. This will be discussed in more details in section \ref{sec:reduction}.

This non-renormalizable Lagrangian describes a low energy effective theory for tensor mesons dynamics. Therefore, its domain of validity is restricted to a certain energy domain which is of the order of the mass of $f_{2}$ mesons because it is the only scale appearing in the problem. Although this cannot be proven rigorously, the results obtained from this theory should be taken with caution at very high transverse momenta.  
 
In high-energy proton-proton collisions, it is well-known from parton distribution functions that gluons dominate the cross-section, so the main contribution to $f_{2}$ production comes from the gluonic sector of the theory. The energy-momentum tensor of gluons can be obtained from the Yang-Mills Lagrangian by varying the metric and is given by
 \begin{eqnarray}
\label{en_mom_tensor01}
T^{\mu \nu}_{\rm{gluons}}(x) = \frac{1}{4}g^{\mu \nu} G_{\sigma \rho a}(x) G_{a}^{\sigma \rho}(x) - G_{\sigma a}^{\; \; \mu}(x) G_{a}^{\sigma \nu}(x)
\end{eqnarray}
where
\begin{eqnarray}
\label{eq:field_st}
G_{a}^{\mu \nu}(x) = \partial^{\mu} A_{a}^{\nu}(x) - \partial^{\nu} A_{a}^{\mu}(x) + gf_{abc}A_{b}^{\mu}(x)A_{c}^{\nu}(x)
\end{eqnarray}
is the usual field-strength tensor and $A^{\mu}_{a}$ is the gauge field of gluons with color index $a$. The Feynman rules associated with this Lagrangian are presented in Appendix \ref{feyn_rule}. This effective theory can now be used to compute the cross-section of $f_{2}$ production in high-energy pp collisions.

%%%%%%%%%%%%%%%%%%%%%%%%%%%%%%%%%%%%%%%%%%%%%%%%%%%%%%%%%%%%%%%%%%%%%%%%%%%%%%%%%%%%%%%%%%%%%%%%%%%%%%%%%%%%%%%

\section{$f_{2}$-Meson Production Cross-Section from pQCD}
\label{pQCD_calc}

In this section, the $f_{2}$-meson cross-section in proton-proton collisions is evaluated in the $k_{\perp}$-factorization formalism. This formalism can be used when the collision is semihard, meaning that  $\Lambda_{QCD}^{2} \ll \mu^{2} \ll s $ where $s$ is the squared center of mass energy, $\Lambda_{QCD} \sim 200 \; \mbox{MeV}$ is the usual QCD scale and $\mu$ is the typical parton interaction scale. The parton interaction scale is related to the transverse mass of the produced particles $\mu^{2} \sim M_{\perp}^{2} \equiv M^{2} + k_{\perp}^{2}$.  For relatively low transverse momenta as the ones measured in experiments (for example, for momenta measured up to $ |k_{\perp}| \lesssim 10 \; \mbox{GeV}$), the interaction scale for $f_{2}$ production is within $M^{2} < \mu^{2} \lesssim 100 \; \mbox{GeV}^{2}$, which obey the inequality for semihard processes.  Clearly, the $f_{2}$ observed at RHIC and LHC should be produced from these semihard processes, unless they are measured at transverse momenta of the order of the center of mass energy ($|k_{\perp}| \sim \sqrt{s}$). Moreover, these processes occur at small-x ($x \approx M_{\perp}^{2}/s \ll 1$) where the collinear factorized perturbation theory is spoiled because of logarithmic terms like $\left[ \ln(\mu^{2}/\Lambda_{QCD}^{2}) \alpha_{s} \right]^{n}$, $\left[ \ln(\mu^{2}/\Lambda_{QCD}^{2}) \ln(1/x) \alpha_{s} \right]^{n}$ and $\left[  \ln(1/x) \alpha_{s} \right]^{n}$ \cite{Collins:1991ty,Catani:1990eg,Gribov:1984tu,Kuraev:1977fs} . The essence of $k_{\perp}$-factorization is to resum these ``large-log", leading to a description in terms of the unintegrated gluon distribution functions $\phi(x,k_{\perp},\mu^2)$. These functions give the probability to find a gluon with longitudinal momentum $x$ and transverse momentum $k_{\perp}$ at the scale $\mu^{2}$ \cite{Collins:1991ty,Catani:1990eg}. 

It should be noted that $k_{\perp}$-factorization have not been proven rigorously for $f_{2}$ production. However, it is analogous to proven heavy quark production described in \cite{Collins:1991ty,Catani:1990eg} and to Higgs production studied in \cite{Lipatov:2005at}. Moreover, as shown in section \ref{sec:prod_CGC}, the Color Glass Condensate can be used to justify this approach since the two formalisms give the same cross-section. This is similar to the work described in \cite{Gelis:2003vh} where it is shown how to recover $k_{\perp}$-factorization from the CGC formalism. For these reasons, using  $k_{\perp}$-factorization for $f_{2}$-production at RHIC and LHC energies is justified.

\subsection{Cross-Section in $k_{\perp}$-factorization}

The inclusive cross-section in the $k_{\perp}$-factorization formalism is given by \cite{Lipatov:2001ny,Zotov:2003cb,Lipatov:2005at}
\begin{eqnarray}
\label{eq:k_cross_section}
(2\pi)^{3}2E_{k}\frac{d\sigma^{pp \rightarrow f_{2} X}}{d^{3}k} &=& 16 \pi^{2} \int_{0}^{1}\frac{dx_{1}}{x_{1}}\frac{dx_{2}}{x_{2}}\int \frac{d^{2}q_{\perp} d^{2}p_{\perp}}{(2\pi)^{4}} \phi_{1}(x_{1},p_{\perp}^{2},\mu^2) \phi_{2}(x_{2},q_{\perp}^{2},\mu^2) \nonumber \\
&& \times (2\pi)^{3}2E_{k}\frac{d\sigma^{g^{*}g^{*} \rightarrow f_{2} X}}{d^{3}k}
\end{eqnarray}
where
\begin{eqnarray}
(2\pi)^{3}2E_{k}\frac{d\sigma^{g^{*}g^{*} \rightarrow f_{2}}}{d^{3}k} = \frac{1}{2\hat{s}} |\mathcal{M}^{g^{*}g^{*} \rightarrow f_{2}}|^{2} (2\pi)^{4} \delta^{4}(p+q-k)
\end{eqnarray}
is the high-energy limit of the cross-section for off-shell gluons $g^{*}$ to on-shell $f_{2}$, $\hat{s} = x_{1}x_{2}s$ is the $k_{\perp}$-factorization flux factor \cite{Collins:1991ty,Catani:1990eg}, $x_{1,2}$ are momentum fractions of gluons and $\phi_{1,2}(x_{1,2},k_{\perp},\mu^2)$ are unintegrated gluon distribution functions of proton 1 and 2. The unintegrated distribution functions are related to usual parton distribution functions of gluons (appearing in collinear factorization ) by
\begin{eqnarray}
\label{eq:k_vs_coll}
\int_{0}^{\mu^2} dk_{\perp}^{2} \phi(x,k_{\perp}^{2},\mu^2) \approx xG(x,\mu^2)
\end{eqnarray}
where $G(x,\mu^2)$ is the usual gluon distribution function in collinear factorization. 

\subsection{Kinematics and Matrix Element}

To compute the production cross-section of $f_{2}$-mesons, the high energy limit of the lowest order matrix element between two off-shell gluons and one on-shell $f_{2}$ has to be calculated. The Feynman diagram included is shown in Fig. (\ref{fig:feynm_f2_k_fac}). It can be evaluated by using the Feynman rules presented in Appendix \ref{feyn_rule}. The main difference with usual pQCD calculation is in the average on polarization of off-shell gluons. In $k_{\perp}$-factorization, the polarization tensor obeys \cite{Collins:1991ty,Catani:1990eg}  
\begin{eqnarray}
\sum_{\lambda} \epsilon^{*\mu}_{\lambda}(p) \epsilon^{\nu}_{\lambda}(p) = \frac{p^{\mu}_{\perp}p^{\nu}_{\perp}}{p_{\perp}^{2}}
\end{eqnarray}
where $p_{\perp}^{\mu} \equiv (0,p_{\perp},0)$. The sum on polarizations differs from the usual result because we are considering off-shell gluons with a virtuality given by $p^{2}=-p_{\perp}^{2}$. The exact form is due to the coupling of gluons to partons through eikonal vertices as well as gauge invariance and Ward identities \cite{Catani:1990eg}.  This can also be shown to be gauge invariant \cite{Catani:1990eg}.
\begin{figure}
\centering
		\includegraphics{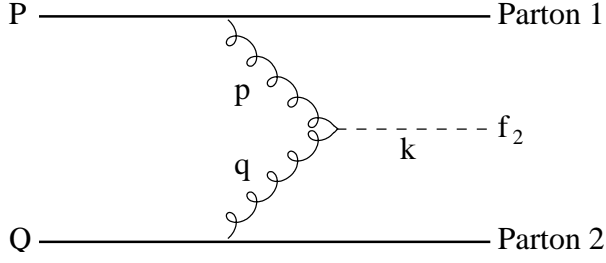}
	\caption{Feynman diagram included in the lowest order calculation of $f_{2}$ production in the $k_{\perp}$-factorization formalism. Two partons with momenta $P$ and $Q$, each part of proton 1 and 2, emit two off-shell gluons with momenta $p$ and $q$. The gluons combine to produce a $f_{2}$ meson with momentum $k$.}
	\label{fig:feynm_f2_k_fac}
\end{figure}

In the center of mass frame, the 4-momenta of partons inside the proton moving in the $\pm z$ direction in Minkowski coordinates can be written,as:
\begin{eqnarray}
P &=& \left(\frac{\sqrt{s}}{2},0,0, \frac{\sqrt{s}}{2}\right) \; ; \;
Q = \left(\frac{\sqrt{s}}{2},0,0, -\frac{\sqrt{s}}{2}\right)
\end{eqnarray}
Then, the momenta of gluons in the large energy limit ($|p_{\perp}|,|q_{\perp}| \ll \sqrt{s}$) is simply 
\begin{eqnarray}
\label{eq:momenta}
p &=& \left(\frac{x_{1}\sqrt{s}}{2},p_{\perp}, \frac{x_{1}\sqrt{s}}{2}\right) \; ; \;
q = \left(\frac{x_{2}\sqrt{s}}{2},q_{\perp}, -\frac{x_{2}\sqrt{s}}{2}\right)
\end{eqnarray}
The $f_{2}$-meson is on-shell and obeys $k^2=M^2$.

The matrix element for $f_{2}$ production is
\begin{eqnarray}
|\mathcal{M}^{g^{*}g^{*} \rightarrow f_{2}}|^{2} & \equiv & \frac{1}{(N_{c}^{2}-1)^2}\sum_{a,b} \sum_{\lambda, \lambda', \lambda''}|\mathcal{T}^{g^{*} g^{*} \rightarrow f_{2}}|^{2} \nonumber \\
 &=& \sum_{a,b} \frac{1}{(N_{c}^{2}-1)^{2} } P_{\mu \nu \alpha \beta}(k) \frac{p_{\perp \rho}p_{\perp \eta}}{p_{\perp}^{2}}  \frac{q_{\perp \sigma}q_{\perp \gamma}}{q_{\perp}^{2}} \nonumber \\
 && \times V^{\mu \nu \rho \sigma}_{ab}(k,q,p) \left[ V^{\alpha \beta \eta \gamma}_{ba}(k,q,p) \right]^{*} 
\end{eqnarray}
where $a,b$ are color indices, $N_{c}$ is the number of color, $ V^{\alpha \beta \eta \gamma}_{ba}(k,q,p)$ are vertices defined in Appendix \ref{feyn_rule} and the projection operator is defined as
\be
 P_{\mu\nu\rho\sigma}(k)
 \equiv
 \sum_{\lambda} 
 \left(\epsilon_{\mu\nu}^\lambda(k)\right)^*\, 
 \epsilon_{\rho\sigma}^\lambda(k) 
 = 
 \left(
 \frac{1}{2}\hat{g}_{\mu\rho} \hat{g}_{\nu\sigma}
 +
 \frac{1}{2} \hat{g}_{\mu\sigma} \hat{g}_{\nu\rho}
 -
 \frac{1}{3}\hat{g}_{\mu\nu} \hat{g}_{\rho\sigma}
 \right)_k
 \label{eq:Pmnrs}
 \ee
 with $\hat{g}_{\mu\nu} = g_{\mu\nu} - {k_\mu k_\nu /M^2}$.

By contracting the tensors $p_{\perp \rho},p_{\perp \eta}$ and $q_{\perp \sigma},q_{\perp \gamma}$ with the vertices, the expression of the matrix element can be simplified further (with $p^{+}$ and $q^{-}$ in light-cone coordinates):
\begin{eqnarray}
|\mathcal{M}^{g^{*}g^{*} \rightarrow f_{2}}|^{2} &=& \frac{(p^{+}q^{-})^{2}}{(N_{c}^{2}-1) \kappa^{2}} P_{\mu \nu \alpha \beta}(k)  \frac{H_{\perp}^{\mu \nu}(p_{\perp},q_{\perp})H_{\perp}^{\rho \sigma}(p_{\perp},q_{\perp})}{p_{\perp}^{2}q_{\perp}^{2}}  
\end{eqnarray}
where we defined $H_{\perp}^{\mu \nu}(p_{\perp},q_{\perp})$ as
\begin{eqnarray}
H_{\perp}^{11}(p_{\perp},q_{\perp}) &=& -H_{\perp}^{22}(p_{\perp},q_{\perp}) = q^{1}p^{1}-q^{2}p^{2} \\
H_{\perp}^{12}(p_{\perp},q_{\perp}) &=& H_{\perp}^{21}(p_{\perp},q_{\perp}) = q^{1}p^{2} + q^{2}p^{1}  
\end{eqnarray}
All the other components of $H_{\perp}^{\mu \nu}$ are zero.

\subsection{Cross-Section}

Once the matrix element is known, it is possible to evaluate the differential cross-section. Using the kinematics (Eq. (\ref{eq:momenta})) and the delta functions, the integrations on $x_{1,2}$ can be easily done. The $k_{\perp}$-factorized differential cross-section is 
\begin{eqnarray}
\label{eq:k_fac_cross}
(2\pi)^{3} 2E_{k}\frac{d\sigma^{pp \rightarrow f_{2} X}}{d^{3}k} &=& 16 \pi^{4} \frac{P_{\mu \nu \alpha \beta}(k)}{(N_{c}^{2}-1) \kappa^{2}}  \int \frac{d^{2}q_{\perp} d^{2}p_{\perp}}{(2\pi)^{4}} (2\pi)^{2} \delta^{2}(p_{\perp}+q_{\perp}-k_{\perp})  \nonumber \\
&& \times \phi_{1}(x_{+},p_{\perp}^{2},\mu^2) \phi_{2}(x_{-},q_{\perp}^{2},\mu^2)    \frac{H_{\perp}^{\mu \nu}(p_{\perp},q_{\perp})H_{\perp}^{\alpha \beta}(p_{\perp},q_{\perp})}{p_{\perp}^{2}q_{\perp}^{2}} 
\end{eqnarray}
where $x_{\pm} = \frac{1}{\sqrt{s}}\left[E_{k} \pm k_{z}  \right]$. This is one of the more important result of this paper. It relates the production cross-section of $f_{2}$ mesons to unintegrated gluon distribution functions. The phenomenology of this equation and its derivation from the CGC formalism are studied in the next sections.

\subsection{Limit of Collinear Factorization}

The procedure to recover collinear factorization cross-sections from $k_{\perp}$-factorization is well-known \cite{Collins:1991ty,Lipatov:2001ny,Zotov:2003cb,Gelis:2003vh} and will serve as a consistency check for Eq. (\ref{eq:k_fac_cross}). The limit $|p_{\perp}|,|q_{\perp}| \rightarrow 0$ has to be taken in the matrix elements and the integration on the azimuthal angle has to be performed. The last step is to use the relation Eq. (\ref{eq:k_vs_coll}) to make the last integral and relate the unintegrated distributions to the collinear distributions.  

For the case of $f_{2}$ production, the integration on azimuthal angles of the $p_{\perp},q_{\perp}$-dependent part of the matrix element is
\begin{eqnarray}
\int_{0}^{2\pi} d\theta_{p} d\theta_{k} \lim_{|p_{\perp}|,|q_{\perp}| \rightarrow 0} P_{\mu \nu \alpha \beta}(k)\frac{H_{\perp}^{\mu \nu}(p_{\perp},q_{\perp})H_{\perp}^{\alpha \beta}(p_{\perp},q_{\perp})}{p_{\perp}^{2}q_{\perp}^{2}} = 8\pi^{2}
\end{eqnarray}
Combining this result with Eq. (\ref{eq:k_vs_coll}) and Eq. (\ref{eq:k_fac_cross}), the cross-section becomes ($s$ is the center of mass energy)
\begin{eqnarray}
\label{eq:cross_coll}
(2\pi)^{3}2E_{k}\frac{d\sigma_{coll.}^{pp \rightarrow f_{2} X}}{d^{3}k} &=&  \frac{2\pi^{2} M^{2}}{(N_{c}^{2}-1) \kappa^{2}s}  (2\pi)^{2} \delta^{2}(k_{\perp})   G(x_{+},\mu^2) G(x_{-},\mu^2)     
\end{eqnarray}
which is exactly the same as the expression computed in the collinear formalism in Appendix \ref{app:coll_cross} (see Eq. \ref{eq:cross_coll02}). Thus, Eq. (\ref{eq:k_fac_cross}) has the right collinear limit.

%%%%%%%%%%%%%%%%%%%%%%%%%%%%%%%%%%%%%%%%%%%%%%%%%%%%%%%%%%%%%%%%%%%%%%%%%%%%%%%%%%%%%%%%%%%%%%%%%%%%%%%%%%%%%%%%

\section{Production of $f_{2}$-Mesons from the CGC}
\label{sec:prod_CGC}

In collisions at very high energy, the wave function of nuclei is dominated by soft gluons (small-$x$ where $x$ is the momentum fraction). The Color Glass Condensate (CGC) is a formalism that describes the dynamics of these degrees of freedom in the limit of large center of mass energy. In this approach, hard partons that carry most of the longitudinal momentum, and soft gluons that have small longitudinal momentum component, are treated differently. Hard partons act as sources for soft gluons and are no longer dynamical degrees of freedom. The occupation number of soft gluons is large because of an emission enhancement at small-x, so classical field equations can be used to understand their dynamics (for reviews of CGC, see \cite{Iancu:2002xk,Iancu:2003xm,Venugopalan:2004dj}). 

In the CGC, computing a physical quantity involves two main steps. The first one is to solve the Yang-Mills equation of motion
\begin{eqnarray}
[D_{\mu},F^{\mu \nu}(x)] = J^{\nu}(x)
\end{eqnarray} 
where the current $J^{\nu}_{a}(x) = \delta^{\nu +} \delta(x^{-})\rho_{1,a}(x_{\perp}) + \delta^{\nu -} \delta(x^{+})\rho_{2,a}(x_{\perp})$ represents random static sources localized on the light-cone \cite{Iancu:2002xk,Iancu:2003xm}. In that case, $\rho_{1,2}(x_{\perp})$ are color charge densities in the transverse plane of proton 1 and 2 respectively. The next step is to take the average over the distribution of color charge densities in the nuclei with weight functionals $W_{1,2}[\rho_{1,2}]$ that include the dynamics of the sources. For any operator that can be related to color charge densities, this can be written as
\begin{eqnarray}
\label{eq:average}
\langle O \rangle = \int \mathcal{D}\rho_{1} \mathcal{D} \rho_{2}  O[\rho_{1},\rho_{2}]W_{1}[x_{1},\rho_{1}]W_{2}[x_{2},\rho_{2}]
\end{eqnarray} 
Computing the weight functional is a highly non-perturbative procedure so it usually involves approximation based on physical modelling. In the limit of a large nuclei at not too small x, it can be approximated by the McLerran-Venugopalan (MV) model, which assumes that the partons are independent sources of color charge \cite{McLerran:1993ka,McLerran:1993ni}. In pp collision, this kind of model fails and a more phenomenological approach, through unintegrated distribution function, is better suited. 

One important ingredient is missing for the computation of $f_{2}$-meson production cross-section. It is necessary to have a relation between the cross-section and a correlator that can be evaluated using Eq. (\ref{eq:average}). This is done in the next section by deriving a reduction formula for our effective theory.

\subsection{Reduction Formula for the Effective Theory}
\label{sec:reduction}

The production of tensor mesons from the CGC can be calculated from a reduction formula derived from the Lagrangian  Eq. (\ref{eq:L_KK}). The first step is to compute the equation of motion of the tensor field $f_{\mu \nu}$ given as usually by $L_{\mu \nu}(x) \equiv \frac{\delta S}{\delta f^{\mu \nu}(x)}=0$. Using $L_{\mu \nu}$ as well as $L_{\mu}^{\mu}$, $\partial^{\mu}L_{\mu \nu}$ and $\partial^{\mu} \partial^{\nu}L_{\mu \nu}$ (see \cite{Giudice:1999ck} for details), the following equation of motion is obtained (note that for the simplicity of this derivation, we chose only the coupling to one type of particles so that the interaction Lagrangian looks like $\frac{1}{\kappa}f_{\mu \nu}(x)T^{\mu \nu}(x)$, the case with the Lagrangian given in Eq. (\ref{eq:int_lag}) can be easily dealt with at the end):
 \be
 (\partial^2 + M^2) f_{\mu\nu} (x)
 =
 \frac{1}{ \kappa} \Theta_{\mu\nu} (x)
 \label{eq:eom_f_mu_nu}
 \ee
 where
 \be
 \Theta_{\mu\nu} (x)
 =
 T_{\mu\nu} (x)
 -
 \left(
 \frac{\partial_\mu\partial_\nu}{ M^2}
 +
 g_{\mu\nu}
 \right)
 \frac{T_\rho^\rho (x)}{ 3} 
 \label{eq:Theta_mu_nu}
 \ee
 as well as the following two constraints:
 \be
 \partial^\mu f_{\mu\nu} (x) = -\frac{\partial_\nu T_\rho^\rho (x)}{3M^2 \kappa}  \;\; \mbox{and} \;\; f_\rho^\rho (x) = -\frac{T_\rho^\rho (x)}{ 3M^2 \kappa} 
 \label{eq:trace_cond}
 \ee
 These constraints make $f_{\mu\nu}$ to have
 the right free-field limit with only 5 propagating  modes. The source term $\Theta_{\mu\nu}$ satisfies
 \be
 \Theta_\rho^\rho (x)
 =
 -\frac{1}{ 3M^2}(\partial^2 + M^2) T_\rho^\rho (x)
 \label{eq:trace_Theta}
 \ee
 and
 \be
 \partial^\mu \Theta_{\mu\nu} (x)
 =
 -\frac{1}{ 3M^2}(\partial^2 + M^2) \partial_\nu T_\rho^\rho (x)
 \label{eq:div_Theta}
 \ee
 The above equations can also be interpreted as operator equations.
 Since the equation of motion is linear in $f_{\mu\nu}$, the general solution of the
 equation of motion Eq. (\ref{eq:eom_f_mu_nu})  is easy to obtain.
 The Fourier transform of the space coordinates to momentum-space is taken and
 the solution for the field operator 
 is given by
 \be
 \hat{f}_{\mu\nu}(t, {\bf k})
 =
 \hat{f}^{(0)}_{\mu\nu}(t, {\bf k})
 +
 \frac{1}{\kappa}\int_{-\infty}^\infty dt'\, G_{\rm ret}(t-t', {\bf k})\, 
 \hat{\Theta}_{\mu\nu}(t', {\bf k})
 \label{eq:f2_sol}
 \ee
 where $\hat{f}^{(0)}_{\mu\nu}(t, {\bf x})$ 
 satisfies the free field equation and the
 retarded Green function in the mixed $t$ and ${\bf k}$ representation is 
% \be
% G_{\rm ret}(x) =
% \int {d^4 k\over(2\pi)^4}\,
% {e^{-ikx}\over -k^2 + M^2 - i\epsilon k^0}
% \ee
 \be
 G_{\rm ret}(t,{\bf k})
 =
 \frac{\sin(E_k t)}{ E_k} \, \theta(t)
 \label{eq:mixedG}
 \ee
It can be easily shown in momentum space that the solution of $f_{\mu \nu}$ obeys the constraints in Eq. (\ref{eq:trace_cond}). Therefore, this solution obeys the requirements of spin-2 particles and describes the right dynamics.

The inclusive cross-section is related to the average number $\bar{n}$ of $f_{2}$ produced. In terms of creation and  annihilation operators, $\bar{n}$ is given by
 \be
 (2\pi)^3 2 E_k \frac{d\bar{n}}{ d^3 k}
 =
 \sum_\lambda
 \bra{\rm init} 
 (\hat{a}_{\bf k}^{\lambda\dagger})_{\rm out} 
 (\hat{a}_{\bf k}^{\lambda})_{\rm out} 
 \ket{\rm init}
 \label{eq:reduction_N_f}
 \ee
where $\lambda = 0, \pm 1, \pm 2$ are the polarization of the spin-2 tensor field.  Here $\ket{\rm init}$ specifies a quantum state at $t = -\infty$ and the subscript `out' specifies that the operator is to be evaluated at  $t = \infty$. As $t\to \infty$, we expect that the density of particles will become low and the fields become asymptotic. We start by writing the creation/annhilation operators in term of the field like
 \be
 (\hat{a}_{\bf k}^\lambda)_{\rm out}
 =
 \lim_{t\to \infty}
 (\epsilon_{\mu\nu}^\lambda(k))^*
 \left[ 
 E_k \hat{f}^{\mu\nu}(t, {\bf k}) 
 + i\partial_t \hat{f}^{\mu\nu}(t, {\bf k}) 
 \right]
 \ee
where the polarization tensor $\epsilon_{\mu\nu}^\lambda$ is traceless, transverse and satisfies
 \be
 (\epsilon_{\mu\nu}^\lambda(k))^*\, 
 \epsilon^{\mu\nu}_{\lambda'}(k) = \delta_{\lambda,\lambda'}
 \ee
 Using the solution \eqn{eq:f2_sol}, we then get
 \be
 (\hat{a}_{\bf k}^\lambda)_{\rm out}
% & = &
% \lim_{t\to \infty}
% \left(\epsilon_{\mu\nu}^\lambda(k)\right)^*
% \left[ 
% E_k \hat{f}^{(0)\mu\nu}(t, {\bf k}) 
% + i\partial_t \hat{f}^{(0)\mu\nu}(t, {\bf k}) 
% \right] 
% \non & & {}
% +
% \lim_{t\to \infty}
% \int_{-\infty}^\infty dt'\, 
% \left[ 
% i\partial_t G_{\rm ret}(t-t',{\bf k})\, 
% +
% E_k G_{\rm ret}(t-t',{\bf k})\, 
% \right]
% \left(\epsilon_{\mu\nu}^\lambda(k)\right)^*
% \tilde{\Theta}_{\mu\nu}(t', {\bf k})
% \non
 &=& 
 \lim_{t\to \infty}
 (\hat{a}_{\bf k}^\lambda e^{-iE_kt})_{\rm free} \nonumber \\
 &&+
 \lim_{t\to \infty}
 \frac{i}{\kappa}
 \int_{-\infty}^t dt'\, 
 e^{- i E_k (t-t')} 
 \left(\epsilon_{\mu\nu}^\lambda(k)\right)^*
 \hat{\Theta}_{\mu\nu}(t', {\bf k})
 \label{eq:reduction}
 \ee
 where we used \eqn{eq:mixedG} to get 
 \be
 i\dot{G}_{\rm ret}(t,{\bf k}) + E_k G_{\rm ret}(t,{\bf k})
% & = &
% i\cos(E_k t) \theta(t)
% + \sin(E_k t) \theta(t)
% + i{\sin(E_k t)\over E_k} \delta(t) 
% \non
 = 
 i e^{- i E_k t} \theta(t)
 \ee
 Here,
 \be
 (\hat{a}_{\bf k}^\lambda e^{-iE_kt})_{\rm free}
 =
 (\epsilon_{\mu\nu}^\lambda(k))^*
 \left[ 
 E_k \hat{f}^{(0)\mu\nu}(t, {\bf k}) 
 + i\partial_t \hat{f}^{(0)\mu\nu}(t, {\bf k}) 
 \right]
 \ee
describes the time evolution of the initial tensor meson population. \eqn{eq:reduction} of course is a reduction formula for $f_{2}$ production. To simplify the analysis, it is assumed that the initial state contains no or little tensor mesons 
 \be
 (\hat{a}_{\bf k}^\lambda)_{\rm free}\,
 \ket{\rm init} 
 = 0 
 \ee
which is not an unreasonable assumption for high energy hadronic collisions. The average number of $f_{2}$ produced is then simply
 \be
 (2\pi)^3 2E_k \frac{d\bar{n}}{ d^3 k}
 =
  \frac{1}{\kappa^{2}}P^{\mu\nu\rho\sigma}(k)
 \langle
 \hat{\Theta}_{\mu\nu}^\dagger(k)
 \hat{\Theta}_{\rho\sigma}(k)
 \rangle
 \label{eq:pre_main}
 \ee
where $\Theta_{\rho\sigma}(k)$ is the Fourier transform of $\Theta_{\rho\sigma}(x)$ evaluated at a tensor meson on-shell momentum. The projection operator $P^{\mu\nu\rho\sigma}(k)$ is obtained by summing over polarizations and is given by Eq. (\ref{eq:Pmnrs}). The angular brackets $\langle \hat{O} \rangle $ here indicates expectation value of $\hat{O}$ in the initial state. Due to the presence of the projection operator in \eqn{eq:pre_main}, some more simplifications can be made.  The partial derivatives in $\Theta_{\mu\nu}$ simply results in factors of $k_\mu k_\nu$ in the Fourier transform. This implies that in the expression (\ref{eq:pre_main}), those partial derivative terms do not contribute due to the transversality of the projection operator ($P^{\mu\nu\rho\sigma}(k) k_\mu = 0$) when $k$ is an on-shell momentum. Furthermore, since $P^{\mu\nu\rho\sigma}$ is made up of $\epsilon_{\alpha\beta}^\lambda$, it is also traceless in the sense that
 $ 
 P^{\mu\nu\rho\sigma}g_{\mu\nu} = 
 P^{\mu\nu\rho\sigma}g_{\rho\sigma} = 0 
 $.
 Hence the $g_{\mu\nu}$ term in $\tilde\Theta_{\mu\nu}$ 
 does not contribute to this momentum distribution, either. By making these simplifications and by doing the same derivation with the interaction Lagrangian in Eq. (\ref{eq:int_lag}), the following simple result is obtained:
 \be
 (2\pi)^3 2E_k \frac{d\bar{n}}{d^3 k}
 =
 \sum_{i,j} \frac{1}{\kappa_{i}\kappa_{j}}P^{\mu\nu\rho\sigma}(k)
 \langle
 \hat{T}_{i,\mu\nu}^\dagger(k)
 \hat{T}_{j,\rho\sigma}(k)
 \rangle
 \label{eq:pre_main02}
 \ee
where the indices are $i,j=\{\rm{gluons},\rm{quarks},\rm{hadrons} \}$. This is the main result of this section. It relates the average number of $f_{2}$ mesons produced to a correlator of energy-momentum tensors. This correlator can then be evaluated using any analytical or numerical methods available. The average $\langle ... \rangle$ depends on the system studied. Looking at a plasma at equilibrium, it could be computed using finite temperature field theory or even ADS/CFT. However, as stated in the introduction, tensor mesons cannot be observed if the thermalized medium lifetime is longer than the decay width of $f_{2}$ mesons. Therefore, these techniques are not pursued here. In high energy hadronic collision, the correlator can be evaluated using the CGC once $T^{\mu \nu}$ is expressed in terms of the sources $\rho_{1,2}$. This is done in the next section.

The result obtained so far is exact given the Lagrangian (\ref{eq:L_KK}). The only assumptions made along the way are:
\begin{enumerate}
	\item There is no tensor mesons in the initial state
	\item The tensor meson content of $T^{\mu\nu}$ is ignored ($T^{\mu \nu}_{f_{2}} \approx 0$)
	\item The tensor meson can be produced on mass shell
\end{enumerate}
The first assumption is justified by the fact that in high-energy collisions, the number of tensor mesons in a nuclei before the collision is negligible. The second assumption is more subtle as it does not appear explicitly in this calculation. If this assumption is not made, the energy-momentum tensor depends on $f_{\mu \nu}$ through the hadronic sector so the equation of motion becomes non-linear and much harder to solve. In a high-energy collision, the cross-section is dominated by gluons, so it is consistent to assume that $T^{\mu \nu}_{f_{2}} \ll T^{\mu \nu}_{\rm{gluons}}$ as long as the $f_{2}$ mesons are produced in a small number. Finally, the last assumptions is that tensor meson can be produced on-shell so that their spectral density is $\rho(M^{2}) \sim \delta(p^{2}-M^{2})$. However, $f_{2}$-mesons are resonances, so the spectral density should look rather as a Breit-Wigner function $\rho(M^{2}) \sim \Gamma/((p^{2} - M^{2})^{2} + M^{2}\Gamma^{2})$ where $\Gamma$ is the decay width. Since $M^{2} \gg \Gamma^{2}$, we expect that corrections due to the finite decay width can be neglected and that we can approximate $\rho(M^{2}) \sim \Gamma/((p^{2} - M^{2})^{2} +  M^{2} \Gamma^{2}) \approx \delta(p^{2}-M^{2})$.  

Having expressed the average number of $f_{2}$ produced in terms of a correlator of energy-momentum tensors, it is possible to compute the cross-section in the CGC formalism. In high-energy collisions, we have that $T^{\mu \nu}_{\rm{gluons}} \gg T^{\mu \nu}_{\rm{quarks}},T^{\mu \nu}_{\rm{hadrons}}$ so we can neglect the quark and hadron contributions. Thus, in the CGC, the inclusive cross-section for $f_{2}$ production is given by \cite{Gelis:2003vh,Baltz:2001dp}
 \be
 (2\pi)^3 2E_k \frac{d\sigma}{ d^3 k}&=& \int d^{2}b_{\perp}(2\pi)^3 2E_k \frac{d\bar{n}(b_{\perp})}{ d^3 k} \nonumber \\
 &=&
 \frac{1}{\kappa^{2}} P_{\mu\nu\rho\sigma}(k) \int d^{2}b_{\perp}
 \int \mathcal{D} \rho_{1} \mathcal{D} \rho_{2}
 \hat{T}^{\dagger \mu\nu}_{\rm{gluons}} [\rho_{1},\rho_{2}]
 \hat{T}^{\rho\sigma}_{\rm{gluons}}[\rho_{1},\rho_{2}] \nonumber \\
 && \times
 W_{1}[\rho_{1}]W_{2}[\rho_{2};b_{\perp}]
 \label{eq:cross_CGC_1}
 \ee
In this expression, $b_{\perp}$ is the impact parameter. The energy-momentum tensor $T^{\mu \nu}_{\rm{gluons}}$ is a functional of the source once the Yang-Mills equation of motion of the gauge field is solved (see Eq. (\ref{en_mom_tensor01}) and (\ref{eq:field_st}) for the expression of the energy-momentum tensor as a function of the gauge field).

\subsection{Calculation of the Cross-Section and relation to $k_{\perp}$-factorization}

In this section, Eq. (\ref{eq:cross_CGC_1}) is evaluated explicitly to leading order for pp collisions. The expansion parameters in a dilute system like a proton are the weak color charge densities obeying $\rho_{1,2}/k_{\perp}^{2} \ll 1$. The first step is to solve the Yang-Mills equation. In a full solution, the gauge field would be a functional to all orders in the color charge densities. However, because the sources are weak (or equivalently, if the typical transverse momentum is large \cite{Venugopalan:2004dj,Gelis:2004bz}), the solution can be truncated and the calculation can be done analytically. The solution of Yang-Mills equation to first order in $\rho_{1,2}/k_{\perp}^{2}$ was found in covariant gauge ($\partial_{\mu}A^{\mu}_{a}(x)=0$) in \cite{Gelis:2003vh,Kovchegov:1997ke}. It is given by $A^{\mu}(k) = A^{\mu}_{1}(k)+A^{\mu}_{2}(k)+A^{\mu}_{12}(k)$ where $A^{\mu}_{1}(k),A^{\mu}_{2}(k)$ are $O(\rho_{1})$ and $O(\rho_{2})$ respectively,  but $A^{\mu}_{12}(k)$ is $O(\rho_{1}\rho_{2})$. The explicit components of the gauge field are 
\begin{eqnarray}
\label{eq:gauge_plus}
A^{+}_{1,a}(k)&=& 2\pi  \delta(k^{-}) \frac{\rho_{1,a}(k_{\perp})}{k_{\perp}^{2}} \\
\label{eq:gauge_minus}
A^{-}_{2,a}(k)&=& 2\pi  \delta(k^{+}) \frac{\rho_{2,a}(k_{\perp})}{k_{\perp}^{2}}
\end{eqnarray}
All the other components are zero. The field $A^{\mu}_{12}$ gives higher order contribution to the energy-momentum tensor correlator $\langle T^{\dagger \mu \nu}(k)  T^{\rho \sigma}(k) \rangle$, so it is not considered here. The leading order contribution to the energy-momentum tensor correlator is $O(\rho^{2}_{1} \rho^{2}_{2})$ and that can only come from different combinations of $A_{1}^{\mu}$ and $A_{2}^{\mu}$. A term containing $A^{\mu}_{12}$ is at least $O(\rho^{3}_{1} \rho^{2}_{2})$ or $O(\rho^{2}_{1} \rho^{3}_{2})$. In principle, terms of $O(\rho^{4}_{1} )$, $O(\rho^{4}_{2} )$, $O(\rho_{1} \rho^{3}_{2} )$ and $O(\rho^{3}_{1} \rho_{2} )$  can also be included but it can be shown that they do not contribute to the cross-section. Finally, the non-Abelian part of the energy-momentum tensor can also be neglected because it gives contributions of at least $O(\rho^{3}_{1} \rho^{2}_{2})$ or $O(\rho^{2}_{1} \rho^{3}_{2})$. Basically, the leading contribution of the energy-momentum tensor correlator in a dilute system is given by the Abelian part of the energy-momentum tensor computed with the gauge field $A_{\rm{leading}}^{\mu} = A_{1}^{\mu}+A_{2}^{\mu}$.

The Abelian part of the energy-momentum tensor in momentum space is
\begin{eqnarray}
\label{eq:T_mom}
T^{\mu \nu}(k) &=& \int \frac{d^{4}p}{(2\pi)^{4}} \biggl\{ \frac{1}{2}g^{\mu \nu} \left[ -p_{\alpha}(k-p)^{\alpha} A_{\beta,a}(p)A^{\beta}_{a}(k-p) + p_{\alpha}(k-p)^{\beta} A_{\beta,a}(p)A^{\alpha}_{a}(k-p)\right] \nonumber \\
&& + p_{\alpha}(k-p)^{\alpha} A^{\mu}_{a}(p)A^{\nu}_{a}(k-p) - p_{\alpha}(k-p)^{\nu} A^{\mu}_{a}(p)A^{\alpha}_{a}(k-p) \nonumber \\
&& -p^{\mu}(k-p)^{\alpha} A_{\alpha,a}(p)A^{\nu}_{a}(k-p) +p^{\mu}(k-p)^{\nu} A_{\alpha,a}(p)A^{\alpha}_{a}(k-p) \biggr\}
\end{eqnarray}
According to our previous discussion, the correlator $\langle T^{\dagger \mu \nu }(k) T^{\rho \sigma }(k) \rangle$ at leading order is given by changing $A^{\mu} \rightarrow A^{\mu}_{\rm{leading}} = A_{1}^{\mu} + A_{2}^{\mu} $ in the expression of the energy-momentum tensor. Then, the computation involves the evaluation of correlators with four gauge fields that have the following general form
\begin{eqnarray}
 \int \frac{d^{4}p d^{4}q}{(2\pi)^{8}} f_{\mu ' \nu ' \rho ' \sigma '}^{{\mu  \nu  \rho  \sigma}}(p,k,q)\langle A^{\dagger \mu'}_{s_{1},a}(p) A^{\dagger \nu'}_{s_{2},a}(k-p)A^{\rho'}_{s_{3},b}(q) A^{\sigma'}_{s_{4},b}(k-q)\rangle
\end{eqnarray}
where $s_{i} = 1 \; \mbox{or} \; 2$ and  $f_{\mu ' \nu ' \rho ' \sigma '}^{{\mu  \nu  \rho  \sigma}}(p,k,q)$ is a function whose value is determined by the Lorentz indices structure and the momentum factors appearing in front of the fields inside the expression of $T^{\mu \nu}(k)$. We need to evaluate the correlator
\begin{eqnarray}
A_{s_{1},s_{2},s_{3},s_{4}}^{\mu  \nu  \rho  \sigma}(k,p,q) \equiv \langle A^{\dagger \mu}_{s_{1},a}(p) A^{\dagger \nu}_{s_{2},a}(k-p)A^{\rho}_{s_{3},b}(q) A^{\sigma}_{s_{4},b}(k-q)\rangle 
\end{eqnarray}
for all values of $s_{i}$ using the expressions of the gauge field in Eqs. (\ref{eq:gauge_plus}) and (\ref{eq:gauge_minus}). They are given by
\begin{eqnarray}
A_{1,2,1,2}^{\mu  \nu  \rho  \sigma}(k,p,q)  &=& 
\delta^{\mu +} \delta^{\nu -} \delta^{\rho +} \delta^{\sigma -} (2\pi)^{4}  \delta(p^{-}) \delta(k^{+} - p^{+}) \delta(q^{-}) \delta(k^{+} - q^{+}) \nonumber \\
&& \times \frac{\langle \rho_{1,a}^{\dagger}(p_{\perp}) \rho_{1,b}(q_{\perp}) \rangle}{p_{\perp}^{2}q_{\perp}^{2}}   \frac{ \langle \rho_{2,a}^{\dagger}(k_{\perp} - p_{\perp}) \rho_{2,b}(k_{\perp} - q_{\perp}) \rangle}{(k_{\perp}-p_{\perp})^{2} (k_{\perp}-q_{\perp})^{2}}
\end{eqnarray}
\begin{eqnarray}
A_{1,2,2,1}^{\mu  \nu  \rho  \sigma}(k,p,q)  &=& 
\delta^{\mu +} \delta^{\nu -} \delta^{\rho -} \delta^{\sigma +} (2\pi)^{4}  \delta(p^{-}) \delta(k^{+} - p^{+}) \delta(q^{+}) \delta(k^{-} - q^{-}) \nonumber \\
&& \times \frac{\langle \rho_{1,a}^{\dagger}(p_{\perp}) \rho_{1,b}(k_{\perp} -q_{\perp}) \rangle }{p_{\perp}^{2} (k_{\perp}-q_{\perp})^{2}}  \frac{\langle \rho_{2,a}^{\dagger}(k_{\perp} - p_{\perp}) \rho_{2,b}( q_{\perp}) \rangle}{q_{\perp}^{2} (k_{\perp}-p_{\perp})^{2}} 
\end{eqnarray}
\begin{eqnarray}
A_{2,1,2,1}^{\mu  \nu  \rho  \sigma}(k,p,q) &=& 
\delta^{\mu -} \delta^{\nu +} \delta^{\rho -} \delta^{\sigma +} (2\pi)^{4}  \delta(p^{+}) \delta(k^{-} - p^{-}) \delta(q^{+}) \delta(k^{-} - q^{-}) \nonumber \\
&& \times \frac{\langle \rho_{1,a}^{\dagger}(k_{\perp}-p_{\perp}) \rho_{1,b}(k_{\perp} -q_{\perp}) \rangle }{(k_{\perp}-p_{\perp})^{2} (k_{\perp}-q_{\perp})^{2}}  \frac{\langle \rho_{2,a}^{\dagger}( p_{\perp}) \rho_{2,b}( q_{\perp}) \rangle}{p_{\perp}^{2}q_{\perp}^{2}} 
\end{eqnarray}
\begin{eqnarray}
A_{2,1,1,2}^{\mu  \nu  \rho  \sigma}(k,p,q) &=& 
\delta^{\mu -} \delta^{\nu +} \delta^{\rho +} \delta^{\sigma -} (2\pi)^{4}  \delta(p^{+}) \delta(k^{-} - p^{-}) \delta(q^{-}) \delta(k^{+} - q^{+}) \nonumber \\
&& \times \frac{\langle \rho_{1,a}^{\dagger}(k_{\perp}-p_{\perp}) \rho_{1,b}(q_{\perp}) \rangle }{q_{\perp}^{2}(k_{\perp}-p_{\perp})^{2} }  \frac{\langle \rho_{2,a}^{\dagger}( p_{\perp}) \rho_{2,b}(k_{\perp} - q_{\perp}) \rangle}{p_{\perp}^{2}(k_{\perp}-q_{\perp})^{2}}
\end{eqnarray}
The other possibilities like $A_{1,1,2,2}$ and $A_{2,2,1,1}$ do not contribute because they give terms proportional to $\delta(k^{\pm})\delta^{2}(k_{\perp})$. Because $k^{2}=M^{2}$, these delta functions have no support. From these expressions, we see clearly that the cross-section is related to averages on sources like Eq. (\ref{eq:average}). The averages can then be evaluated using any model available. However, it is convenient for phenomenological applications to first relate averages to unintegrated gluon distribution functions.

These averages can be related to the unintegrated distribution function of $k_{\perp}$-factorization like \cite{Blaizot:2004wu,Blaizot:2004wv} 
\begin{eqnarray}
\label{eq:coor_unint1}
 \langle \rho_{1,a}^{\dagger}(p_{\perp}) \rho_{1,b}(q_{\perp}) \rangle &=&\frac{4\pi^{2} \delta_{ab}}{ (N^{2}_{c}-1)}\left[\frac{p_{\perp}+q_{\perp}}{2} \right]^{2}  \nonumber \\
&& \times \int d^{2}x_{\perp} e^{i(p_{\perp}-q_{\perp})\cdot x_{\perp}} \frac{d\phi_{1}\left(\frac{p_{\perp}+q_{\perp}}{2} | x_{\perp}\right)}{d^{2}x_{\perp}} \\
\label{eq:coor_unint2}
 \langle \rho_{2,a}^{\dagger}(p_{\perp}) \rho_{2,b}(q_{\perp}) \rangle &=&\frac{4\pi^{2} \delta_{ab}}{ (N^{2}_{c}-1)}\left[\frac{p_{\perp}+q_{\perp}}{2} \right]^{2} \nonumber \\
&& \times \int d^{2}y_{\perp} e^{i(p_{\perp}-q_{\perp})\cdot (y_{\perp}+  b_{\perp})} \frac{d\phi_{2}\left(\frac{p_{\perp}+q_{\perp}}{2} | y_{\perp} \right)}{d^{2}y_{\perp}}
\end{eqnarray}
where $\frac{d\phi_{1,2}\left(p_{\perp} | y_{\perp} \right)}{d^{2}y_{\perp}}$ are the unintegrated gluon distribution functions per unit area. By definition, they are related to the unintegrated distribution functions by 
\begin{eqnarray}
\phi_{1,2}\left(p_{\perp} \right) = \int d^{2}y_{\perp} \frac{d\phi_{1,2}\left(p_{\perp} | y_{\perp} \right)}{d^{2}y_{\perp}}
\end{eqnarray}
where the integration is on the transverse extent of the nuclei. 

Combining all these results together, it is possible to compute the cross-section. The integration on the impact factor $\int d^{2} b_{\perp}$ in the definition of the cross-section gives a delta function $(2\pi)^{2} \delta^{2}(p_{\perp}-q_{\perp})$ in the second source average Eq. (\ref{eq:coor_unint2}). After a lengthy but straightforward calculation, we find that the cross-section is:
\begin{eqnarray}
\label{eq:crossCGC}
(2\pi)^{3}2E_{k}\frac{d\sigma^{pp \rightarrow f_{2} X}}{d^{3}k} &=&  16\pi^{4} \frac{P_{\mu \nu \alpha \beta}(k)}{(N_{c}^{2}-1) \kappa^{2}} \int \frac{d^{2}q_{\perp} d^{2}p_{\perp}}{(2\pi)^{4}} (2\pi)^{2} \delta^{2}(p_{\perp}+q_{\perp}-k_{\perp})  \nonumber \\
&& \times \phi_{1}(p_{\perp}^{2},\mu^2) \phi_{2}(q_{\perp}^{2},\mu^2)    \frac{H_{\perp}^{\mu \nu}(p_{\perp},q_{\perp})H_{\perp}^{\alpha \beta}(p_{\perp},q_{\perp})}{p_{\perp}^{2}q_{\perp}^{2}} 
\end{eqnarray}
This has exactly the same structure as the cross-section obtained in $k_{\perp}$-factorization shown in Eq. (\ref{eq:k_fac_cross}) but now,it is derived as the low density limit of CGC. The longitudinal momentum fraction $x$ dependence of the unintegrated distribution functions is introduced through quantum evolution.  The weight functions $W_{1,2}[\rho_{1,2}]$ obeys a non-linear evolution equation in $x$, the JIMWLK equation \cite{Iancu:2000hn,Ferreiro:2001qy,Jalilian-Marian:1996xn,Jalilian-Marian:1997gr}. This makes the unintegrated gluon distribution function $x$-dependent $\phi(q_{\perp}^{2},\mu^2) \rightarrow \phi(x,q_{\perp}^{2},\mu^2)$ and Eqs. (\ref{eq:crossCGC}) and (\ref{eq:k_fac_cross}) can be compared.

\section{Phenomenology for Proton-Proton Collisions}

Eq. (\ref{eq:crossCGC}) (or (\ref{eq:k_fac_cross})) can be used to study the phenomenology of $f_{2}$ production and to study different parameterizations of the unintegrated gluon distribution function. By contracting the indices of $P_{\mu \nu \alpha \beta}(k)H_{\perp}^{\mu \nu}(p_{\perp},q_{\perp})H_{\perp}^{\alpha \beta}(p_{\perp},q_{\perp}) $, the cross-section at mid-rapidity is given by
\begin{eqnarray}
\label{eq:cross_mid01}
\left. \frac{d\sigma^{pp \rightarrow f_{2} X}}{d^{2}k_{\perp}dy} \right|_{y=0} &=&  \frac{1}{2\pi(N_{c}^{2}-1) \kappa^{2}} \int d^{2}q_{\perp} d^{2}p_{\perp}    \phi_{1}(x',p_{\perp}^{2},\mu^2) \phi_{2}(x',q_{\perp}^{2},\mu^2) \nonumber \\
&& \times \delta^{2}(k_{\perp} -p_{\perp} - q_{\perp}) \nonumber \\
&& \times \biggl\{ 1+ \frac{k_{\perp}^{2}}{M^{2}} +  \frac{\left[ p_{\perp}^{2} (q_{\perp} \cdot k_{\perp})  +q_{\perp}^{2} (p_{\perp} \cdot k_{\perp}) \right]^{2}}{3M^{4}p_{\perp}^{2}q_{\perp}^{2}}  \biggr\}
\end{eqnarray}
where $x'=\sqrt{\frac{M^{2}+k_{\perp}^{2}}{s}}$. As shown in this expression, the precise measurement of the $f_{2}$ differential cross-section is a direct probe of unintegrated distribution functions.  The unintegrated distribution functions obey evolution equations like the BFKL or the CCFM equation. Depending on the physical model used or the approximations involved in the solution of these equations, the unintegrated distribution function can be parametrized in various ways. To get an estimate of $f_{2}$ production that can be compared with experimental data at RHIC, we use standard parameterizations \cite{Andersson:2002cf}\footnote{We would like to thanks H. Jung for handing us his FORTRAN routine CAUNIGLU which evaluates numerically all of these parameterizations. It can be found at http://www.desy.de/~jung/cascade/updf.html.}. The measurement of $f_{2}$ production can be used to put constraints on models of the unintegrated gluon distribution function by comparing the predictions of many approaches. In this paper, we use the following parametrizations.

\subsection{DIG (Derivative of the Integrated Gluon distribution function)}

By ignoring that the unintegrated distribution function depends on a factorization scale $\mu$, it is possible to invert Eq. (\ref{eq:k_vs_coll}) to get
\begin{eqnarray}
\phi_{DIG}(x,p_{\perp}^{2}) = \left. \frac{d xg(x,\mu^{2})}{d\mu^{2}} \right|_{\mu^{2}=p_{\perp}^{2}}
\end{eqnarray}
where $xg(x,\mu^{2})$ is the gluon parton distribution function. 

\subsection{JB (J. Blumlein)}

The JB parameterization is based on a perturbative solution of the BFKL equation \cite{Andersson:2002cf,Blumlein:1995eu}. In this parameterization, the unintegrated gluon distribution function is written as
\begin{eqnarray}
 \phi_{JB}(x,p_{\perp}^{2},\mu^2) = 
\int_{x}^{1}dz \mathcal{G}(z,p_{\perp},\mu^2) \frac{x}{z} G\left(\frac{x}{z},\mu^{2} \right)
\end{eqnarray}
where $G(x,\mu^{2})$ is the collinear gluon distribution function and 
\begin{eqnarray}
 \mathcal{G}(z,p_{\perp}^{2},\mu^2)= \left\{ 
\begin{array}{cc}
\frac{\bar{\alpha}_{s}}{ z p_{\perp}^{2}}J_{0}\left( \sqrt{\bar{\alpha}_{s} \ln\left( \frac{1}{z} \right) \ln \left( \frac{\mu^{2}}{p_{\perp}^{2}} \right)} \right),&\mbox{ if } p_{\perp}^{2} \leq \mu^{2}\\
\frac{\bar{\alpha}_{s}}{ z p_{\perp}^{2}}I_{0}\left( \sqrt{\bar{\alpha}_{s} \ln\left( \frac{1}{z} \right) \ln \left( \frac{p_{\perp}^{2}}{\mu^{2}} \right)} \right),&\mbox{ if } p_{\perp}^{2} > \mu^{2}
\end{array}\right.
\end{eqnarray}
In this last equation, $J_{0}$ and $I_{0}$ are the Bessel's and modified Bessel's functions of the first kind  respectively and $\bar{\alpha}_{s} = 3 \alpha_{s}/\pi$ where $\alpha_{s}$ is the strong coupling constant ($\alpha_{s} \approx 0.25$).

\subsection{CCFM (Catani,Ciafaloni,Fiorani,Marchesini)}

The unintegrated gluon distribution function have been calculated numerically by solving the CCFM \cite{Ciafaloni:1987ur,Catani:1989sg,Catani:1989yc}
 evolution equations using a Monte Carlo method \cite{Jung:2000hk}. The initial conditions of the evolution equation are determined from a fit of the proton structure function $F_{2}(x,Q^{2})$. The result of this procedure for many data sets is implemented in the routine CAUNIGLU written by H. Jung.

\subsection{KMR (Kimber,Martin, Ryskin)}

The KMR parameterization starts at a certain scale given by $p_{\perp 0}^{2} \sim 1 \; \mbox{GeV}^{2}$. The non-perturbative part is given by the MRST collinear distribution function. The unintegrated distribution function is given by \cite{Andersson:2002cf,Kimber:2001sc}
\begin{eqnarray}
 \phi_{KMR}(x,p_{\perp}^{2},\mu^{2})= \left\{ 
\begin{array}{cc}
\frac{xg(x,p_{\perp 0}^{2})}{p_{\perp 0}^{2}} & \mbox{ if } p_{\perp}^{2} < p_{\perp 0}^{2}\\
\frac{\mathcal{G}_{KMR}(x,p_{\perp}^{2},\mu^{2})}{p_{\perp}^{2}} &\mbox{ if } p_{\perp}^{2} \geq p_{\perp 0}^{2}
\end{array}\right.
\end{eqnarray}
where the function $\mathcal{G}_{KMR}(x,p_{\perp}^{2},\mu^{2})$ now depends on the scale $\mu$ and can be evaluated numerically.

\begin{figure}
\begin{center}
\includegraphics[width=0.7\textwidth]{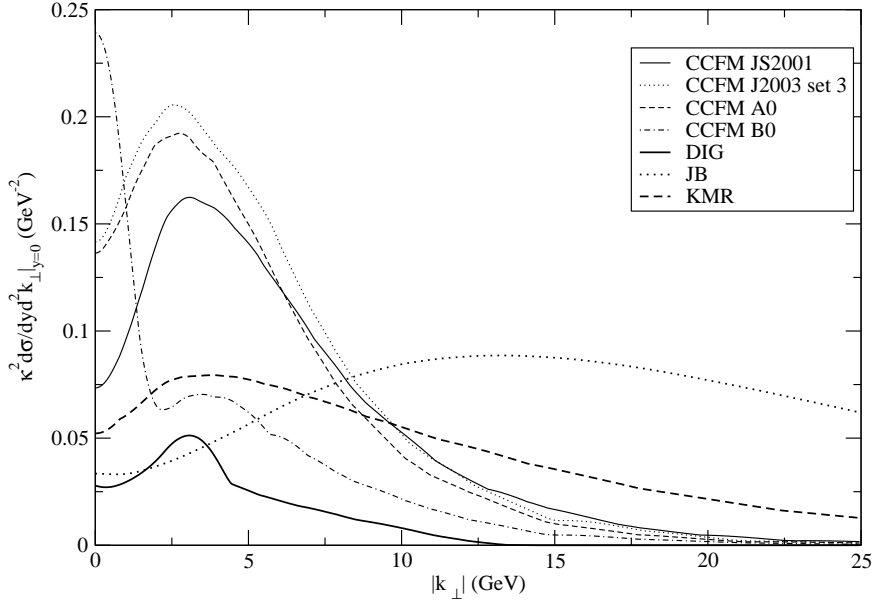}
\end{center}
\caption{Numerical results of the differential cross-section as a function of transverse momentum $|k_{\perp}|$ for different parameterizations of the unintegrated distribution function at RHIC energy ($\sqrt{s}=200 \; \mbox{GeV}$). The results are shown for midrapidity ($y=0$).}
\label{fig:result_RHIC_ds}
\end{figure}

\begin{figure}
\begin{center}
\includegraphics[width=0.7\textwidth]{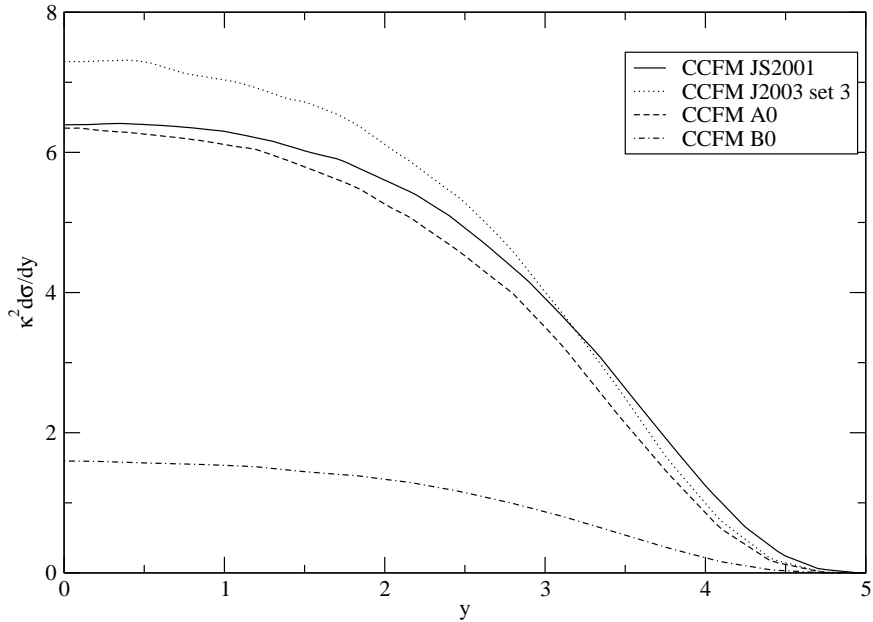}
\end{center}
\caption{Numerical results of the differential cross-section as a function of rapidity $y$ for the CCFM parameterizations of the unintegrated distribution function at RHIC energy ($\sqrt{s}=200 \; \mbox{GeV}$). The transverse momentum $k_{\perp}$ has been integrated.}
\label{fig:result_totalRHIC_ds}
\end{figure}

For numerical computations, we set the number of color to $N_{c} = 3$, the center of mass energy to $\sqrt{s} \approx 200 \; \mbox{GeV}$ (RHIC) and the mass of $f_{2}$ to $1.27 \; \mbox{GeV}$. For the DIG parametrization, we use the GRV NLN collinear gluon distribution function and for the JB parametrization, we use the MRS collinear distribution. The numerical results of the differential cross-section at midrapidity for seven different parameterizations at RHIC energy are presented in Fig. \ref{fig:result_RHIC_ds}. In Fig. \ref{fig:result_totalRHIC_ds}, we present the results for the $k_{\perp}$-integrated (for $0<|k_{\perp}|<\sqrt{s}$) differential cross-section for the CCFM parameterizations. All the numerical integrations are done with the CUBA package \cite{Hahn:2004fe} using both CUHRE and VEGAS algorithms. 

There are some qualitative differences between the predictions of the different unintegrated distribution functions. As shown in Fig. \ref{fig:result_RHIC_ds}, the CCFM parameterizations are decreasing much faster at large $|k_{\perp}|$ compared to other parameterizations. At low $|k_{\perp}|$, the shape of the curves is different, especially the parameterization CCFM B0.  These features give us a way to discriminate between the unintegrated distribution functions. As seen in Fig. \ref{fig:result_totalRHIC_ds}, the parameterizations also lead to results that have different magnitudes. However, there is still an uncertainty on the overall magnitude because of the value of the coupling constant that have to be fixed by experiments. We now discuss this crucial point.

It is possible to compare our predictions to experimental data to fix the value of coupling constant $\kappa$. The STAR collaboration have measured the production of $f_{2}$ mesons in the invariant mass spectrum of pions in proton-proton collisions at $\sqrt{s} \approx 200 \; \mbox{GeV}$ and midrapidity \cite{Adams:2003cc}. The present data only show the invariant mass distribution in the transverse momentum bin $0.6 \; \mbox{GeV} < |k_{\perp}|< 0.8 \; \mbox{GeV}$. From the graph presented in \cite{Adams:2003cc}, it is possible to extract a rough estimate of the average number of $f_{2}$ produced per collision. Taking into account that the branching ratio of $f_{2} \rightarrow \pi \pi$ is $\rm{BR}_{f_{2} \rightarrow \pi \pi} \approx 0.85$ \cite{Yao:2006px}, the average number of $f_{2}$ produced per collision is $\tilde{n} = \int_{-0.5}^{0.5}dy \int_{0.6 \; GeV}^{0.8 \; GeV}  d^{2}k_{\perp} \frac{d\bar{n}}{d^{2}k_{\perp}dy}  \approx 0.0024$. The total inelastic cross-section for pp-collisions at $\sqrt{s} = 200 \; \mbox{GeV}$ is $\sigma_{\rm{inel}} \approx 40 \; \mbox{mb}$. Thus, we can convert the average multiplicity to a cross-section and we get that 
\begin{eqnarray}
\tilde{\sigma} \approx \sigma_{\rm{inel}} \int_{-0.5}^{0.5}dy \int_{0.6 \; GeV}^{0.8 \; GeV}  d^{2}k_{\perp} \frac{d\bar{n}}{d^{2}k_{\perp}dy}  \approx 0.095 \; \mbox{mb} \; \; \; \; (\mbox{Experimental})
\end{eqnarray}
This is an estimation of the experimental result for the $f_{2}$ production cross-section.

We can then compute the integrated cross-section $\tilde{\sigma}$ using Eq. (\ref{eq:k_fac_cross}). The results are presented in Table \ref{table:results}. This can then be used to determine the value of the coupling constant $\kappa$ and we find that depending on the parametrization used, $\kappa$ is between $0.415$ GeV and $0.850$ GeV. This is of the same order of magnitude but much larger that $\kappa \approx 0.1$ GeV obtained from pion decay by assuming a universal coupling of $f_{2}$ to the energy momentum tensor of all strongly interacting particles \cite{suzuki,Katz:2005ir} or by tensor meson dominance \cite{Bongardt:1980qf}. The coupling of $f_{2}$ to gluons is much weaker than the coupling to pions. There are two reasons that explain this discrepancy. First, we are assuming that the $f_{2}$ meson can be produced as a stable on-shell particle. In $f_{2}$ decay, it is obviously not the case, the $f_{2}$ is unstable and decays in pions (or any decay product). Second, there are no justification \textit{a priori} that the partonic sector should have the same coupling constant as that of the hadronic sector. In that sense, it is expected that the coupling of $f_{2}$ to gluons and hadrons will be different. Once the coupling constant is determined, all the curves presented in Fig. \ref{fig:result_RHIC_ds} and  \ref{fig:result_totalRHIC_ds} become predictions and can be compared to experimental data.

\begin{table}
\begin{tabular}{|c|c|c|}
\hline
Parameterization	 & $\kappa^{2} \tilde{\sigma}  $  & $\kappa$ (GeV) \\
\hline \hline
 CCFM JS2001 & 0.0767 &  0.558  \\
 CCFM J2003 set 3& 0.142  & 0.759 \\
 CCFM A0 & 0.134   & 0.737  \\
 CCFM B0 & 0.178   & 0.850  \\
 DIG & 0.0424  & 0.415 \\
 JB& 0.0578 &  0.484\\
 KMR &  0.0708  & 0.536  \\
\hline
\end{tabular}
\caption{Numerical result for the total integrated cross-section using different parameterization of the unintegrated distribution functions. To compare with STAR data, we integrated over $0.6\; \mbox{GeV}<|k_{\perp}|<0.8 \; \mbox{GeV}$ and $-0.5<y<0.5$. From this result, it is possible to fix the value of the coupling constant $\kappa$. }
\label{table:results}
\end{table}

\section{Conclusion}

Using our effective theory based on a coupling between the $f_{2}$ meson and the energy-momentum tensor, we were able to compute the inclusive cross-section of $f_{2}$ tensor mesons in proton-proton collisions. This was done in the $k_{\perp}$-factorization formalism and in the CGC formalism. We showed that in the low density limit where the saturation effects are not taken into account, the result of the calculation of the cross-section in the CGC formalism reduces to the result of $k_{\perp}$-factorization. In some sense, we use the CGC to ``prove" Eq. (\ref{eq:k_cross_section}) for $f_{2}$ production. A rigorous treatment of factorization would involve a separation between the long and short distance dynamics using resummation and a reorganization of the perturbative expansion at all orders. Of course, this is a hard task to perform. In the CGC, the proof reduces solely to a power counting in the color charge densities. This kind of result was expected as it was shown in other explicit calculations, namely for gluon \cite{Gyulassy:1997vt,Kovchegov:1997ke} and heavy quark production \cite{Gelis:2003vh}.   

We also looked into the phenomenology of $f_{2}$-meson production in proton-proton collisions at RHIC. We computed the differential cross-sections for many different parameterizations of the unintegrated distribution function. We used STAR data to fix the coupling constant and found that it is larger than in other approaches, even if it has the same order of magnitude. These numerical results, although they should be supplemented by more experimental data, can be used for the determination of the most accurate unintegrated distribution function. This is very interesting since $k_{\perp}$-factorization is one of the main computationnal tool for particle production (like heavy quarks \cite{Collins:1991ty,Catani:1990eg,Jung:2001rp,Lipatov:2001ny,Zotov:2003cb} and Higgs bosons \cite{Lipatov:2005at}) in high-energy collisions at RHIC,LHC and Tevatron. Having a consistent and precise unintegrated gluon distribution function is a crucial element in this formalism. Up to now, the main processes used to determine this distribution function were heavy quarks, dijets and gauge bosons production in pp and electron-proton collisions \cite{Hansson:2007de}. Many improvements have been performed to fit experimental data from these distributions, but there are still many uncertainties, especially at small $k_{\perp}$. This uncertainty can also be seen in the variability of our numerical results for different parameterizations since they essentially lead to very different predictions. The $f_{2}$ production gives another observable that can be used to put constraints on the value of unintegrated distribution functions. 

Our work is also a starting point for further studies such as the production of $f_{2}$ in proton-nucleus collisions. According to CGC ideas, saturation effects should play an important role in that kind of process and thus, the $f_{2}$ could be used to look at initial state effects. In that sense, our calculation can serve as a point of comparison for the presence and the magnitude of these initial state effects. This is under investigation and should be the topics of a future publication.

\begin{acknowledgments}

The authors want to thank F. Gelis, R. Venugopalan, T. Lappi, K. Tuchin and J.-S. Gagnon for interesting and stimulating discussions. We would like to thank especially Y. Kovchegov for clearing out some important aspects of the calculation. We also thank H. Jung for his help with the unintegrated distribution functions. This work was supported  by the Natural Sciences and Engineering Research Council of Canada.

\end{acknowledgments}

%%%%%%%%%%%%%%%%%%%%%%%%%%%%%%%%%%%%%%%%%%%%%%%%%%%%%%%%%%%%%%%%%%%%%%%%%%%%%%%%%%%%%%%%%%%%%%%%%%%%%%%%%%%%%%%%%%%

\appendix
%\section{}

\section{Feynman rules}
\label{feyn_rule}

In this Appendix, the Feynman rules for gluon-$f_{2}$ interaction are presented. By referring to the interaction Lagrangian Eqs. (\ref{eq:int_lag}),(\ref{en_mom_tensor01} and (\ref{eq:field_st}), it is easily seen that there are three possible interaction vertices: $gg \rightarrow f_{2}$, $ggg \rightarrow f_{2}$ and $gggg \rightarrow f_{2}$. However, the second and third ones are $O(g)$ and $O(g^{2})$ respectively. They are not part of the leading order contribution, so they are not considered in this analysis. The vertex $gg \rightarrow f_{2}$ can be evaluated and is given by
\begin{eqnarray}
\label{eq:vertex}
V^{\mu \nu \rho \sigma}_{ab}(k,p,q) &=& \frac{i}{\kappa} (2\pi)^{4}\delta^{4}(-k+p+q) \delta_{ab} \biggl[ g^{\mu \sigma} g^{\nu \rho} (q\cdot p) + g^{\mu \rho} g^{\nu \sigma} (q\cdot p) \nonumber \\
&& - g^{\mu \nu} g^{\sigma \rho} (q\cdot p)   - g^{\mu \sigma} q^{\rho} p^{\nu} - g^{\mu \rho} q^{\nu} p^{\sigma} - g^{\nu \rho} q^{\mu} p^{\sigma}  - g^{\nu \sigma} q^{\rho} p^{\mu} \nonumber \\
&& + g^{\rho \sigma} q^{\mu} p^{\nu}   + g^{\rho \sigma} q^{\nu} p^{\mu}  + g^{\mu \nu} q^{\rho} p^{\sigma} \biggr]
\end{eqnarray}
where $\kappa \approx O(100 \; \mbox{MeV})$ to be fixed by experiments. The vertex has the following important property:
\begin{eqnarray}
\label{prop_vert}
p_{\rho} V^{\mu \nu \rho \sigma}_{ab}(k,p,q) = q_{\sigma}V^{\mu \nu \rho \sigma}_{ab}(k,p,q) = 0
\end{eqnarray}
which is related to the conservation of the Abelian part of the energy-momentum tensor. The external $f_{2}$ have the polarization $\epsilon_{\mu \nu}^{\lambda}(k)$. The $f_{2}$ Feynman propagator is 
\begin{eqnarray}
G_{\mu \nu \rho \sigma}(p) = \frac{-iP_{\mu \nu \rho \sigma}(p)}{p^2 - M^2 + i \epsilon}
\end{eqnarray}
where $P_{\mu \nu \rho \sigma}(p)$ is the projection operator given by Eq. (\ref{eq:Pmnrs}).

\section{Collinear Factorized Cross-Section}
\label{app:coll_cross}

The tensor meson production from pp-collision is investigated in this appendix, using the usual parton model (the collinear factorization formalism). This formalism can be used when $\Lambda_{QCD}^{2} \ll \mu^{2} \sim s $ where again $s$ is the squared center of mass energy, $\Lambda_{QCD} \sim 200 \; \mbox{MeV}$ is the usual QCD scale and $\mu$ is the typical parton interaction scale. At RHIC or LHC energy, this inequality is recovered at very large transverse momentum, when $\mu^{2} \approx M_{\perp}^{2} \sim s$, which may not be physically observable. However, it is interesting to make the calculation using this method as a consistency check for the $k_{\perp}$-factorized formalism.

The cross-section for $f_{2}$-meson production in the collinear factorization formalism is given by
\begin{eqnarray}
(2\pi)^{3}2E_{k}\frac{d\sigma^{pp \rightarrow f_{2}X}}{d^{3}k} &=& \int_{0}^{1}dx_{1} dx_{2} G^{p_{1}}(x_{1},Q^{2}) G^{p_{2}}(x_{2},Q^{2}) \nonumber \\
&& \times (2\pi)^{3} 2E_{k}\frac{d\sigma^{gg \rightarrow f_{2}}}{d^{3}k}
\end{eqnarray}
where $G^{p_{1},p_{2}}$ are the gluon distribution function of the two protons, $x_{1,2}$ are the longitudinal momentum fraction ($x \equiv \frac{p^{+}_{parton}}{p^{+}_{hadron}}$) and $Q^{2}$ is the factorization scale. The cross-section for on-shell gluons to on-shell  tensor meson ($gg \rightarrow f_{2}$) is
\begin{eqnarray}
(2\pi)^{3} 2E_{k}\frac{d\sigma^{gg \rightarrow f_{2}}}{d^{3}k} = \frac{1}{4|p \cdot q|} |\mathcal{M}^{gg \rightarrow f_{2}}|^{2} (2\pi)^{4} \delta^{4}(k-p-q)
\end{eqnarray}
where $\mathcal{M}^{gg \rightarrow f_{2}}$ is the matrix element for the process $gg \rightarrow f_{2}$ (see figure \ref{fig:feynm_f2_k_fac} and take on-shell gluons). The matrix element can be easily computed to lowest order from the Feynman rules described in Appendix \ref{feyn_rule}. 

The matrix element squared is averaged over the degrees of freedom of the initial state and summed over the degrees of freedom of the final state. Once conservation of energy and momentum are used, it is given by
\begin{eqnarray}
|\mathcal{M}^{gg \rightarrow f_{2}}|^{2} &=& \frac{1}{(N_{c}-1)^{2}}\sum_{a,b} \frac{1}{4}\sum_{\lambda, \lambda', \lambda''}|\mathcal{T}^{gg \rightarrow f_{2}}|^{2} = \frac{M^{4}}{2(N_{c}^{2} -1) \kappa^{2}} 
\end{eqnarray}
where $\lambda$, $\lambda'$ and $\lambda''$ are the polarizations of the gluons and $f_{2}$-mesons. This is obtained using the fact that for on-shell gluons, the sum on polarization is
\begin{eqnarray}
\sum_{\lambda} \epsilon_{\mu}(k) \epsilon_{\nu}^{*}(k) &=& -g_{\mu \nu} + \frac{n_{\mu} k_{\nu} + n_{\nu} k_{\mu}}{n \cdot k} + \frac{n^{2}k_{\mu} k_{\nu}}{(n \cdot k)^{2}} \rightarrow -g_{\mu \nu}
\end{eqnarray}
The replacement of the sum over polarizations by $g_{\mu \nu}$ (as depicted in the second part of the equation) is possible only because of the property (\ref{prop_vert}) of the vertex. The sum over polarizations of $f_{2}$-meson is given by Eq. (\ref{eq:Pmnrs}).

The initial gluons do not have transverse momentum in the collinear factorization formalism ($p_{\perp}= q_{\perp} = 0$). Then, the momentum fractions are given by $x_{1} = \frac{2p_{z}}{\sqrt{s}}$ and $x_{2} = \frac{2q_{z}}{\sqrt{s}}$ and we can write the four-momenta as
\begin{eqnarray}
p &=& \left(\frac{x_{1}\sqrt{s}}{2},0,0, \frac{x_{1}\sqrt{s}}{2} \right) \; ; \;
q = \left(\frac{x_{2}\sqrt{s}}{2},0,0, -\frac{x_{2}\sqrt{s}}{2} \right)
\end{eqnarray}
Using the expression for the matrix element and integrating the longitudinal components with the delta functions, the cross-section becomes
\begin{eqnarray}
\label{eq:cross_coll02}
(2\pi)^{3}2E_{k}\frac{d\sigma^{pp \rightarrow f_{2} X}}{d^{3}k} &=&  \frac{2\pi^{2} M^{2}}{(N_{c}^{2}-1) \kappa^{2}s}  (2\pi)^{2} \delta^{2}(k_{\perp})  G(x_{+},Q^2) G(x_{-},Q^2)     
\end{eqnarray}
The exact same equation was obtained by computing the collinear limit of $k_{\perp}$-factorization, Eq. (\ref{eq:cross_coll}). Similar results were obtained for $\eta'$-meson production \cite{Jalilian-Marian:2001bu}.


\begin{thebibliography}{00}


%\cite{Collins:1991ty}
\bibitem{Collins:1991ty}
  J.~C.~Collins and R.~K.~Ellis,
  %``Heavy quark production in very high-energy hadron collisions,''
  Nucl.\ Phys.\  B {\bf 360}, 3 (1991).
  %%CITATION = NUPHA,B360,3;%%

%\cite{Catani:1990eg}
\bibitem{Catani:1990eg}
  S.~Catani, M.~Ciafaloni and F.~Hautmann,
  %``High-energy factorization and small x heavy flavor production,''
  Nucl.\ Phys.\  B {\bf 366}, 135 (1991).
  %%CITATION = NUPHA,B366,135;%%

%\cite{Gribov:1984tu}
\bibitem{Gribov:1984tu}
  L.~V.~Gribov, E.~M.~Levin and M.~G.~Ryskin,
  %``Semihard Processes In QCD,''
  Phys.\ Rept.\  {\bf 100}, 1 (1983).
  %%CITATION = PRPLC,100,1;%%

%\cite{Kuraev:1977fs}
\bibitem{Kuraev:1977fs}
  E.~A.~Kuraev, L.~N.~Lipatov and V.~S.~Fadin,
  %``The Pomeranchuk Singularity In Nonabelian Gauge Theories,''
  Sov.\ Phys.\ JETP {\bf 45}, 199 (1977)
  [Zh.\ Eksp.\ Teor.\ Fiz.\  {\bf 72}, 377 (1977)].
  %%CITATION = ZETFA,72,377;%%


%\cite{Jung:2001rp}
\bibitem{Jung:2001rp}
  H.~Jung,
  %``Heavy quark production at the TEVATRON and HERA using k(t)  factorization
  %with CCFM evolution,''
  Phys.\ Rev.\  D {\bf 65}, 034015 (2002)
  [arXiv:hep-ph/0110034].
  %%CITATION = PHRVA,D65,034015;%%

%\cite{Lipatov:2001ny}
\bibitem{Lipatov:2001ny}
  A.~V.~Lipatov, V.~A.~Saleev and N.~P.~Zotov,
  %``Heavy quark production at the TEVATRON in the semihard QCD approach and
  %the unintegrated gluon distribution,''
  arXiv:hep-ph/0112114.
  %%CITATION = HEP-PH/0112114;%%

%\cite{Zotov:2003cb}
\bibitem{Zotov:2003cb}
  N.~P.~Zotov, A.~V.~Lipatov and V.~A.~Saleev,
  %``Heavy-quark production in p anti-p collisions and unintegrated gluon
  %distributions,''
  Phys.\ Atom.\ Nucl.\  {\bf 66}, 755 (2003)
  [Yad.\ Fiz.\  {\bf 66}, 786 (2003)].
  %%CITATION = YAFIA,66,786;%%
  
%\cite{Lipatov:2005at}
\bibitem{Lipatov:2005at}
  A.~V.~Lipatov and N.~P.~Zotov,
  %``Higgs boson production at hadron colliders in the k(T)-factorization
  %approach,''
  Eur.\ Phys.\ J.\  C {\bf 44}, 559 (2005)
  [arXiv:hep-ph/0501172].
  %%CITATION = EPHJA,C44,559;%%
  
%\cite{Andersson:2002cf}
\bibitem{Andersson:2002cf}
  B.~Andersson {\it et al.}  [Small x Collaboration],
  %``Small x phenomenology: Summary and status,''
  Eur.\ Phys.\ J.\  C {\bf 25}, 77 (2002)
  [arXiv:hep-ph/0204115].
  %%CITATION = EPHJA,C25,77;%%
  
  
  
%%%%%%%%%%%%% Collinear Factorization %%%%%%%%%%%%%%%%%%%%

%\cite{Brock:1993sz}
\bibitem{Brock:1993sz}
  R.~Brock {\it et al.}  [CTEQ Collaboration],
  %``Handbook of perturbative QCD: Version 1.0,''
  Rev.\ Mod.\ Phys.\  {\bf 67}, 157 (1995).
  %%CITATION = RMPHA,67,157;%%

%\cite{Jalilian-Marian:2001bu}
\bibitem{Jalilian-Marian:2001bu}
  J.~Jalilian-Marian and S.~Jeon,
  %``Probing gluons in nuclei: The case of eta',''
  Phys.\ Rev.\  C {\bf 65}, 065201 (2002)
  [arXiv:hep-ph/0110417].
  %%CITATION = PHRVA,C65,065201;%%
  

%%%%%%%%%%%%%%%%%%% CGC %%%%%%%%%%%%%%%

%\cite{McLerran:1993ka}
\bibitem{McLerran:1993ka}
  L.~D.~McLerran and R.~Venugopalan,
  %``Gluon distribution functions for very large nuclei at small transverse
  %momentum,''
  Phys.\ Rev.\  D {\bf 49}, 3352 (1994)
  [arXiv:hep-ph/9311205].
  %%CITATION = PHRVA,D49,3352;%%

%\cite{McLerran:1993ni}
\bibitem{McLerran:1993ni}
  L.~D.~McLerran and R.~Venugopalan,
  %``Computing quark and gluon distribution functions for very large nuclei,''
  Phys.\ Rev.\  D {\bf 49}, 2233 (1994)
  [arXiv:hep-ph/9309289].
  %%CITATION = PHRVA,D49,2233;%%

\bibitem{Iancu:2002xk}
  E.~Iancu, A.~Leonidov and L.~McLerran,
  %``The colour glass condensate: An introduction,''
  arXiv:hep-ph/0202270.
  %%CITATION = HEP-PH 0202270;%%

%\cite{Iancu:2003xm}
\bibitem{Iancu:2003xm}
  E.~Iancu and R.~Venugopalan,
  %``The color glass condensate and high energy scattering in QCD,''
  arXiv:hep-ph/0303204.
  %%CITATION = HEP-PH 0303204;%%

%\cite{Venugopalan:2004dj}
\bibitem{Venugopalan:2004dj}
  R.~Venugopalan,
  %``The color glass condensate: A summary of key ideas and recent
  %developments,''
  arXiv:hep-ph/0412396.
  %%CITATION = HEP-PH 0412396;%%
  
  
%\cite{Gelis:2003vh}
\bibitem{Gelis:2003vh}
  F.~Gelis and R.~Venugopalan,
  %``Large mass q anti-q production from the color glass condensate,''
  Phys.\ Rev.\ D {\bf 69}, 014019 (2004)
  [arXiv:hep-ph/0310090].
  %%CITATION = HEP-PH 0310090;%%

%\cite{Gyulassy:1997vt}
\bibitem{Gyulassy:1997vt}
  M.~Gyulassy and L.~D.~McLerran,
  %``Yang-Mills radiation in ultrarelativistic nuclear collisions,''
  Phys.\ Rev.\ C {\bf 56}, 2219 (1997)
  [arXiv:nucl-th/9704034].
  %%CITATION = NUCL-TH 9704034;%%

%\cite{Kovchegov:1997ke}
\bibitem{Kovchegov:1997ke}
  Y.~V.~Kovchegov and D.~H.~Rischke,
  %``Classical gluon radiation in ultrarelativistic nucleus nucleus
  %collisions,''
  Phys.\ Rev.\ C {\bf 56}, 1084 (1997)
  [arXiv:hep-ph/9704201].
  %%CITATION = HEP-PH 9704201;%%
  

%%%%%%%%%%%%%% Properties of f_2 %%%%%%%%%%%%%%%%%%%

%\cite{Yao:2006px}
\bibitem{Yao:2006px}
  W.~M.~Yao {\it et al.}  [Particle Data Group],
  %``Review of particle physics,''
  J.\ Phys.\ G {\bf 33}, 1 (2006).
  %%CITATION = JPHGB,G33,1;%%
  

%%%%%%%%%%%%%% Chiral Perturbation theory
 
\bibitem{sjrey1}
 {\it Chiral Perturbation Theory for Tensor Mesons}\\
 Chi-Keung Chow and Soo-Jong Rey
 JHEP 9805 (1998) 010
 
 \bibitem{sjrey2}
 {\it Quenched and Partially Quenched Chiral Perturbation Theory for
 Vector and Tensor Mesons}\\
 Chi-Keung Chow and Soo-Jong Rey
 Nucl.Phys. B528 (1998) 303-321
 
 
\bibitem{Toublan:1996bk} 
D.~Toublan,
%``Lowest tensor-meson resonances contributions to the chiral
%perturbation theory low energy coupling constants,''
Phys.\ Rev.\ {\bf D53}, 6602 (1996)
%hep-ph/9509217.
%%CITATION = PHRVA,D53,6602;%%
 
%\cite{Giacosa:2005bw}
\bibitem{Giacosa:2005bw}
  F.~Giacosa, T.~Gutsche, V.~E.~Lyubovitskij and A.~Faessler,
  %``Decays of tensor mesons and the tensor glueball in an effective field
  %approach,''
  Phys.\ Rev.\ D {\bf 72}, 114021 (2005)
  [arXiv:hep-ph/0511171].
  %%CITATION = HEP-PH 0511171;%%



%%%%%%%%%%%%%%%%%%% TMD Refs %%%%%%%%%%%%%%%%%%%%

\bibitem{renner70}
B.~Renner,
Phys.\ Lett.\ B {bf 33}, 599 (1970).

\bibitem{renner71}
B.~Renner,
Nucl.\ Phys.\ B {bf 30}, 634 (1971).

%\cite{Gampp:1978ym}
\bibitem{Gampp:1978ym}
W.~Gampp and H.~Genz,
%``Tensor Meson Dominance And The New Particles,''
Phys.\ Lett.\ B {\bf 76}, 319 (1978).
%%CITATION = PHLTA,B76,319;%%

%\cite{Gampp:1978xc}
\bibitem{Gampp:1978xc}
W.~Gampp and H.~Genz,
%``Absolute Gamma Decay Rates Of The F(C) (3.55) From Tensor Meson Dominance
%And QCD Potentials,''
Phys.\ Lett.\ B {\bf 79}, 267 (1978).
%%CITATION = PHLTA,B79,267;%%

%\cite{Gampp:1979hp}
\bibitem{Gampp:1979hp}
W.~Gampp and H.~Genz,
%``Photonic Decays Involving The 2++ (Anti-B B): Partial Widths And Jets
%From Tensor Meson Dominance,''
Z.\ Phys.\ C {\bf 1}, 199 (1979).
%%CITATION = ZEPYA,C1,199;%%

%\cite{Bongardt:1980qf}
\bibitem{Bongardt:1980qf}
K.~Bongardt, W.~Gampp and H.~Genz,
%``Tensor Meson Dominance, Internal Symmetries And Meson Mixing For The New
%Particles,''
Z.\ Phys.\ C {\bf 3}, 233 (1980).
%%CITATION = ZEPYA,C3,233;%%

%\cite{Genz:1982yn}
\bibitem{Genz:1982yn}
H.~Genz,
%``Tensor Meson Dominance In The Upsilon System,''
Phys.\ Rev.\ D {\bf 26}, 3108 (1982).
%%CITATION = PHRVA,D26,3108;%%

%\cite{Ishikawa:1988xi}
\bibitem{Ishikawa:1988xi}
K.~Ishikawa, I.~Tanaka, K.~Liu and B.~A.~Li,
%``Tensor Meson Dominance And Glueball Candidate Theta (1720),''
Phys.\ Rev.\ D {\bf 37}, 3216 (1988).
%%CITATION = PHRVA,D37,3216;%%

%\cite{Terazawa:1990es}
\bibitem{Terazawa:1990es}
H.~Terazawa,
%``F(2) Dominance Of The Energy Momentum Tensor,''
Phys.\ Lett.\ B {\bf 246}, 503 (1990).
%%CITATION = PHLTA,B246,503;%%
 
\bibitem{suzuki} M.~Suzuki, Phys. Rev. {\bf D47}, 1043 (1993)

%\cite{Yan:1996xq}
\bibitem{Yan:1996xq}
Y.~Yan and R.~Tegen,
%``Role of tensor meson pole and Delta exchange diagrams in p anti-p $\to$
%pi+ pi-,''
Phys.\ Rev.\ C {\bf 54}, 1441 (1996).
%%CITATION = PHRVA,C54,1441;%%


 %\cite{Katz:2005ir}
\bibitem{Katz:2005ir}
  E.~Katz, A.~Lewandowski and M.~D.~Schwartz,
  %``Tensor mesons in AdS/QCD,''
  Phys.\ Rev.\ D {\bf 74}, 086004 (2006)
  [arXiv:hep-ph/0510388].
  %%CITATION = HEP-PH 0510388;%% 
  
  
%%%%%%%%%% Experimental results



%\cite{Adams:2003cc}
\bibitem{Adams:2003cc}
  J.~Adams {\it et al.}  [STAR Collaboration],
  %``rho0 production and possible modification in Au + Au and p + p  collisions
  %at s(NN)**(1/2) = 200-GeV,''
  Phys.\ Rev.\ Lett.\  {\bf 92}, 092301 (2004)
  [arXiv:nucl-ex/0307023].
  %%CITATION = NUCL-EX 0307023;%%   





%%%%%%%%%%%%%%%%%%%%%%%%%%%%%%%%%%%%%%%%%%%%%%%%%%%%%%%

%\cite{Giudice:1999ck}
\bibitem{Giudice:1999ck}
G.~F.~Giudice, R.~Rattazzi and J.~D.~Wells,
%``Quantum gravity and extra dimensions at high-energy colliders,''
Nucl.\ Phys.\ B {\bf 544}, 3 (1999)
[hep-ph/9811291].
%%CITATION = HEP-PH 9811291;%%

%\cite{Han:1999sg}
\bibitem{Han:1999sg}
T.~Han, J.~D.~Lykken and R.~Zhang,
%``On Kaluza-Klein states from large extra dimensions,''
Phys.\ Rev.\ D {\bf 59}, 105006 (1999)
[hep-ph/9811350].
%%CITATION = HEP-PH 9811350;%%


%%%%%%%%%%%%%%%% JIMWLK

%\cite{Iancu:2000hn}
\bibitem{Iancu:2000hn}
  E.~Iancu, A.~Leonidov and L.~D.~McLerran,
  %``Nonlinear gluon evolution in the color glass condensate. I,''
  Nucl.\ Phys.\  A {\bf 692}, 583 (2001)
  [arXiv:hep-ph/0011241].
  %%CITATION = NUPHA,A692,583;%%

%\cite{Ferreiro:2001qy}
\bibitem{Ferreiro:2001qy}
  E.~Ferreiro, E.~Iancu, A.~Leonidov and L.~McLerran,
  %``Nonlinear gluon evolution in the color glass condensate. II,''
  Nucl.\ Phys.\  A {\bf 703}, 489 (2002)
  [arXiv:hep-ph/0109115].
  %%CITATION = NUPHA,A703,489;%%

%\cite{Jalilian-Marian:1996xn}
\bibitem{Jalilian-Marian:1996xn}
  J.~Jalilian-Marian, A.~Kovner, L.~D.~McLerran and H.~Weigert,
  %``The intrinsic glue distribution at very small x,''
  Phys.\ Rev.\  D {\bf 55}, 5414 (1997)
  [arXiv:hep-ph/9606337].
  %%CITATION = PHRVA,D55,5414;%%



%\cite{Jalilian-Marian:1997gr}
\bibitem{Jalilian-Marian:1997gr}
  J.~Jalilian-Marian, A.~Kovner, A.~Leonidov and H.~Weigert,
  %``The Wilson renormalization group for low x physics: Towards the high
  %density regime,''
  Phys.\ Rev.\  D {\bf 59}, 014014 (1999)
  [arXiv:hep-ph/9706377].
  %%CITATION = PHRVA,D59,014014;%%

%\cite{Jalilian-Marian:1997jx}
\bibitem{Jalilian-Marian:1997jx}
  J.~Jalilian-Marian, A.~Kovner, A.~Leonidov and H.~Weigert,
  %``The BFKL equation from the Wilson renormalization group,''
  Nucl.\ Phys.\  B {\bf 504}, 415 (1997)
  [arXiv:hep-ph/9701284].
  %%CITATION = NUPHA,B504,415;%%



%%%%%%%%%%%%% CGC %%%%%%%%%%%%%%%%%%%%%%%%%%%%%%%%%%%

%\cite{Baltz:2001dp}
\bibitem{Baltz:2001dp}
  A.~J.~Baltz, F.~Gelis, L.~D.~McLerran and A.~Peshier,
  %``Coulomb corrections to e+ e- production in ultra-relativistic nuclear
  %collisions,''
  Nucl.\ Phys.\  A {\bf 695}, 395 (2001)
  [arXiv:nucl-th/0101024].
  %%CITATION = NUPHA,A695,395;%%
  
%\cite{Gelis:2004bz}
\bibitem{Gelis:2004bz}
  F.~Gelis and R.~Venugopalan,
  %``Initial state effects in the color glass condensate,''
  J.\ Phys.\ G {\bf 30}, S995 (2004)
  [arXiv:hep-ph/0403229].
  %%CITATION = HEP-PH 0403229;%%


%\cite{Blaizot:2004wu}
\bibitem{Blaizot:2004wu}
  J.~P.~Blaizot, F.~Gelis and R.~Venugopalan,
  %``High energy p A collisions in the color glass condensate approach. I:
  %Gluon production and the Cronin effect,''
  Nucl.\ Phys.\ A {\bf 743}, 13 (2004)
  [arXiv:hep-ph/0402256].
  %%CITATION = HEP-PH 0402256;%%

%\cite{Blaizot:2004wv}
\bibitem{Blaizot:2004wv}
  J.~P.~Blaizot, F.~Gelis and R.~Venugopalan,
  %``High energy p A collisions in the color glass condensate approach. II:
  %Quark production,''
  Nucl.\ Phys.\  A {\bf 743}, 57 (2004)
  [arXiv:hep-ph/0402257].
  %%CITATION = NUPHA,A743,57;%%

% Unintegrated distribution function

%\cite{Blumlein:1995eu}
\bibitem{Blumlein:1995eu}
  J.~Blumlein,
  %``On the k(T) dependent gluon density of the proton,''
  arXiv:hep-ph/9506403.
  %%CITATION = HEP-PH/9506403;%%
  
%\cite{Ryskin:1995wz}
\bibitem{Ryskin:1995wz}
  M.~G.~Ryskin and Yu.~M.~Shabelski,
  %``The Role of screening corrections in small x behavior of structure
  %functions,''
  Z.\ Phys.\  C {\bf 66}, 151 (1995)
  [Phys.\ Atom.\ Nucl.\  {\bf 58}, 1782 (1995\ YAFIA,58,1884-1889.1995)].
  %%CITATION = YAFIA,58,1884;%%

%\cite{Ciafaloni:1987ur}
\bibitem{Ciafaloni:1987ur}
  M.~Ciafaloni,
  %``Coherence Effects in Initial Jets at Small q**2 / s,''
  Nucl.\ Phys.\  B {\bf 296}, 49 (1988).
  %%CITATION = NUPHA,B296,49;%%



%\cite{Catani:1989sg}
\bibitem{Catani:1989sg}
  S.~Catani, F.~Fiorani and G.~Marchesini,
  %``Small X Behavior Of Initial State Radiation In Perturbative QCD,''
  Nucl.\ Phys.\  B {\bf 336}, 18 (1990).
  %%CITATION = NUPHA,B336,18;%%

%\cite{Catani:1989yc}
\bibitem{Catani:1989yc}
  S.~Catani, F.~Fiorani and G.~Marchesini,
  %``QCD Coherence In Initial State Radiation,''
  Phys.\ Lett.\  B {\bf 234}, 339 (1990).
  %%CITATION = PHLTA,B234,339;%%






%\cite{Jung:2000hk}
\bibitem{Jung:2000hk}
  H.~Jung and G.~P.~Salam,
  %``Hadronic final state predictions from CCFM: The hadron-level Monte  Carlo
  %generator CASCADE,''
  Eur.\ Phys.\ J.\  C {\bf 19}, 351 (2001)
  [arXiv:hep-ph/0012143].
  %%CITATION = EPHJA,C19,351;%%

%\cite{Golec-Biernat:1999qd}
\bibitem{Golec-Biernat:1999qd}
  K.~Golec-Biernat and M.~Wusthoff,
  %``Saturation in diffractive deep inelastic scattering,''
  Phys.\ Rev.\  D {\bf 60}, 114023 (1999)
  [arXiv:hep-ph/9903358].
  %%CITATION = PHRVA,D60,114023;%%

%\cite{Kimber:2001sc}
\bibitem{Kimber:2001sc}
  M.~A.~Kimber, A.~D.~Martin and M.~G.~Ryskin,
  %``Unintegrated parton distributions,''
  Phys.\ Rev.\  D {\bf 63}, 114027 (2001)
  [arXiv:hep-ph/0101348].
  %%CITATION = PHRVA,D63,114027;%%

%\cite{Kwiecinski:1997ee}
\bibitem{Kwiecinski:1997ee}
  J.~Kwiecinski, A.~D.~Martin and A.~M.~Stasto,
  %``A unified BFKL and GLAP description of F2 data,''
  Phys.\ Rev.\  D {\bf 56}, 3991 (1997)
  [arXiv:hep-ph/9703445].
  %%CITATION = PHRVA,D56,3991;%%









  
%\cite{Dumitru:2001ux}
\bibitem{Dumitru:2001ux}
  A.~Dumitru and L.~D.~McLerran,
  %``How protons shatter colored glass,''
  Nucl.\ Phys.\ A {\bf 700}, 492 (2002)
  [arXiv:hep-ph/0105268].
  %%CITATION = HEP-PH 0105268;%%
  
%\cite{Gelis:2005pt}
\bibitem{Gelis:2005pt}
  F.~Gelis and Y.~Mehtar-Tani,
  %``Gluon propagation inside a high-energy nucleus,''
  Phys.\ Rev.\ D {\bf 73}, 034019 (2006)
  [arXiv:hep-ph/0512079].
  %%CITATION = HEP-PH 0512079;%%
  
%\cite{Kovchegov:1998bi}
\bibitem{Kovchegov:1998bi}
  Y.~V.~Kovchegov and A.~H.~Mueller,
  %``Gluon production in current nucleus and nucleon nucleus collisions in  a
  %quasi-classical approximation,''
  Nucl.\ Phys.\  B {\bf 529}, 451 (1998)
  [arXiv:hep-ph/9802440].
  %%CITATION = NUPHA,B529,451;%%


%\cite{Hansson:2007de}
\bibitem{Hansson:2007de}
  M.~Hansson and H.~Jung,
  %``Towards precision determination of uPDFs,''
  arXiv:0707.4276 [hep-ph].
  %%CITATION = ARXIV:0707.4276;%%

%\cite{Hahn:2004fe}
\bibitem{Hahn:2004fe}
  T.~Hahn,
  %``CUBA: A library for multidimensional numerical integration,''
  Comput.\ Phys.\ Commun.\  {\bf 168}, 78 (2005)
  [arXiv:hep-ph/0404043].
  %%CITATION = CPHCB,168,78;%%




























\end{thebibliography}
\end{document}